%% file: ECHG24.tex
             \newif\ifsubmission\submissionfalse     

\documentclass[letter,12pt,openany]{article}

\pdfoutput=1

\usepackage[legalpaper, margin=1in]{geometry}
\usepackage{amsmath,amsfonts,amssymb,amsthm}
\usepackage{authblk}
\usepackage{color}
\usepackage{url}
\usepackage{alltt}
\usepackage{ifthen}
\usepackage{svn-multi}
\usepackage[latin1]{inputenc}
\usepackage{times}
\usepackage[T1]{fontenc}
\usepackage[pdftex]{graphicx}
\usepackage{subcaption}
\usepackage[font={small}]{caption}
\usepackage{array}
\usepackage[pdftex,colorlinks]{hyperref}
\graphicspath{{figs/}} 

\input{defsElton}

\title{Chaotic mixing in plane Couette turbulence}

\author[1]{John R. Elton\thanks{jelton.physics@gmail.com}}
\author[1]{Predrag Cvitanovi\'{c}\thanks{predrag.cvitanovic@physics.gatech.edu}}
\author[2]{Jonathan Halcrow\thanks{halcrow@gmail.com}}
\author[3]{John F. Gibson\thanks{john.gibson@unh.edu}}

\affil[1]{School of Physics, Georgia Inst. of Technology, Atlanta GA}
\affil[2]{Google Research, Atlanta, GA}
\affil[3]{Mathematics and Statistics, Univ. New Hampshire, Durham NH}

\begin{document}
    
\date{December 1, 2024}

\maketitle

\begin{abstract}    
Lagrangian tracer particle trajectories for invariant solutions 
of the Navier-Stokes equations confined to the three-dimensional geometry of plane 
Couette flow are studied. Treating the Eulerian 
velocity field of an invariant solution as a dynamical system, the transport of these passive 
scalars along Lagrangian flow trajectories reveals a rich repertoire of 
different types of motion that can occur, including stagnation points, 
for which there is no fluid movement, and invariant tori, which obstruct 
chaotic mixing across the full volume of the plane Couette flow minimal 
cell. We determine the stability of these stagnation points, along with 
their stable and unstable manifolds, and find heteroclinic 
connections between them. These topological features produce a skeleton 
that shapes the passive tracer flow for a turbulent fluid, providing a 
first step to elucidating Lagrangian particle transport and mixing in 
three-dimensional Navier-Stokes turbulent flows. 
\end{abstract}       

\section{Introduction}
\label{s:intro}

The turbulent transport and mixing of different particles or species 
within a fluid is a problem with both wide practical application as well 
as theoretical interest, yet a complete understanding of the phenomena 
remains elusive; even questions related to how we define or measure 
various mixing properties are not universally agreed upon. In 
\cite{MaMPe05}, some pitfalls of standard approaches such as measuring 
variation from homogeneity with an $L^2$ or $L^p$ norm, or computing the 
entropy of the underlying dynamical system, are pointed out. Furthermore, 
there are experimental and computational challenges involved when 
studying the problem in the natural Lagrangian frame 
\cite{MHPRS07,ABBBBB08,BrLiEc06,MoLePi04}. Although the idea of taking a 
dynamical systems approach to the problem is not new, as books by Ottino 
\cite{Botti89} and Wiggins \cite{Wiggins1992} attest to the value of
using invariant manifolds to study fluid transport, \cite{MHPRS07} and 
\cite{Haller02} point out that Lagrangian coherent structures in 
real flow data are difficult to identify due to the uncertain stability 
of individual particles. Thus many of the theoretical and experimental 
analyses are confined to \textit{two-dimensional} systems, with a large 
body of the work on Lagrangian dynamics focusing on the statistical 
properties and fluctuations of particle velocities, and on detecting 
intermittency or anomalous scaling laws 
\cite{EgeChi22,MoLePi04,ABBBBB08,FaGaVe01}. 

In this study, we extend the idea of looking at the Lagrangian transport 
of passive scalars by means of the invariant structures within the flow 
in a \textit{truly $3D$ system}, partitioning the physical space of the 
fluid in a way that reveals distinct types of motion that can occur, 
driving the organization of tracer mixing \cite{Haller02}. By building 
upon the computational work that has provided exact invariant solutions 
of the fully resolved {\NSe}s for {\pCf}, 
described below, we are able to use equilibrium velocity field solutions 
to study a tractable, yet still complex problem that lends itself to a 
dynamical systems analysis. Symmetry considerations allow for a first 
tangible step that will lead to piecing together a full phase portrait of 
such an equilibrium flow, by determining the fixed points and their 
stabilities along with {\hc}s.  Our eventual goal is 
then putting this information together to assist in understanding how to 
calculate quantities to best characterize turbulent fluid mixing. 

The plane Couette geometry we study is a shear flow in which two infinite 
plates move in opposite directions at constant speed, with turbulent 
behavior beginning to set in approximately above Reynolds number $Re=325$ 
\cite{GHCV08}. Eulerian equilibrium velocity fields have been computed 
for this setup over a number of years, and \pCf\ also admits periodic, 
relative periodic, and traveling wave solutions \cite{GHCV08,DV04}.  In 
1990 Nagata \cite{N90} discovered what are known as the upper branch and 
lower branch equilibria by continuing a known solution from 
Taylor-Couette flow to plane Couette. Later, Waleffe \cite{W03} 
calculated the same solutions a different way and noted that they satisfy 
'shift-rotate' and 'shift-reflect' symmetry. Gibson et al. \cite{GHCW07} 
began explorations of plane Couette dynamics around those equilibria, 
making use of the symmetries and noting that the subspace of velocity 
fields under the action of certain symmetry groups was invariant under 
{\NSe}s. The search for new invariant solutions focused on this 
subspace, from which a Newton search was able to detect more equilibria. 
The reader may consult \cite{GHCV08} or \cite{GHCW07} for additional 
history of the computational discoveries of invariant solutions for 
{\pCf}. 

Much of the analysis in this work is carried out on a particular 
equilibrium solution referred to as the "upper branch" or $EQ_2$. We also 
repeat some of our analysis  for another equilibrium velocity field 
$EQ_8$, for which the flow is more turbulent and possesses different 
invariant symmetries. For analyzing fluid particle trajectories from the 
Lagrangian perspective, where we follow the motion of a tracer within a 
fixed equilibrium,  we need to make a distinction between $3D$ physical 
fluid flow for a given invariant solution of {\NSe}s and the dynamical 
$\infty$-dimensional \statesp\ flow. We distinguish between the two by 
using physically motivated nomenclature for the $3D$ physical fluid flow: 
We shall refer to the position for which $\bu(\bx_{_{SP}})=0$ as the {\em 
\stagp} $\bx_{_{SP}}$ or point $SP$. And when we discuss coherent 
structures and {\hc}s, these again refer to trajectories \textit{within} 
a known Eulerian equilibrium velocity field, in contrast to the {\hc}s 
described in, for example \cite{GHCV08}, which track the evolution of the 
velocity fields themselves. 

In \refsect{s:NS}-\refsect{s:PCF_symm} we review the underlying equations 
and geometry for {\pCf}, describe how the equilibria are stored 
numerically for use in computing Lagrangian trajectories, and give a deep 
dive on the symmetries which are crucial for later analysis. Much of the 
information in these sections is a rehash that can be found in other 
places including \cite{GHCW07}, but is important for understanding the 
new contributions of this work.  In \refsect{s:symm_stag} we show how 
the known symmetries automatically provide us with critical information 
for analyzing Lagrangian trajectories within each equilibrium by 
determining where the velocity field must be exactly 0; in other words we 
are able to locate the "fixed points" in dynamical systems terminology, 
or {\stagp}s in our lingo. In \refsect{s:Lagrangian} we give our core 
analysis and results: namely a dynamical systems treatment of Lagrangian 
trajectories within plane Couette equilibria that includes a treatment of 
fixed points, stability analysis and invariant manifolds, and {\hc}s, 
providing the basic dynamical skeleton through which transport and mixing 
properties in a turbulent flow field may be analyzed. We provide an 
intriguing graphical phase portrait of the turbulent motion within the 
upper branch equilibrium and also provide some results for $EQ_8$ and 
discuss potential applications.

\section{Plane Couette Flow}
\label{s:PCF}

\subsection{The Navier-Stokes equations}
\label{s:NS}
 The underlying equations
that govern the motion of \pCf\ are the {\NSe}s,
along with boundary conditions. The boundary conditions for \pCf\ in the $x$
and $z$ directions are periodic,
 $ \bu(x, y, z) = \bu(x+L_x, y, z) =
\bu(x, y, z + L_z) $.
 In the $y$ direction,
 $\bu = (1,0,0)$ at $\bx = (0,1,0)$ and $\bu = (-1,0,0)$ at $\bx =
 (0,-1,0)$.

 The fluid is taken to be incompressible, so in this case the
 {\NSe}s are
 \beq
 \frac{\partial \bu}{\partial t} + (\bu \cdot \nabla)\bu = -\nabla p + \frac{1}{Re} \nabla^{2} \bu
    \,,\qquad
\nabla \cdot \bu  = 0 \,. \label{eqn:NavierStokes} \eeq

For an equilibrium velocity field that is not changing in time, the first equation in
\refeq{eqn:NavierStokes} simplifies to \beq
 (\bu \cdot \nabla)\bu = -\nabla p + \frac{1}{Re} \nabla^{2} \bu
    \,,\qquad \label{eqn:NavierStokes2} \eeq
 
  The Reynolds number parameter $\Reynolds$, which gives a measure of fluid viscosity and degree to which fluid motion may become turbulent, is given by \beq Re = \frac{\overline{u}L}{\nu}
\eeq where $\overline{u}$ is the average fluid velocity and $L$ is
the characteristic length. Thus the form of the {\NSe}s and boundary conditions make use of rescaling to use non-dimensionalized variables. 
We use $\Reynolds = 400$, in the regime of moderate turbulence, for the \pCf\ simulations throughout the text unless otherwise indicated.

For computational purposes, it is easier to work with a velocity field  that
represents the {\em difference} from the laminar flow. 
So we can break up the total field into two components: $\butot =
y \hat{\bf x} + \bu$. Here $y \hat{\bf x}$ is the laminar velocity
field and $\bu$ is then the difference between the total velocity and
laminar. Substitute $y \hat{\bf x} + \bu$ for $\bu$ in the
nondimensionalized {\NSe}s above to get
\beq
    \frac{\partial \bu}{\partial t}
    + y  \frac{\partial \bu}{\partial x}
    + v \, \hat{\bf x}
    + \bu \cdot \bnabla \bu
=
    - \bnabla p
    + \frac{1}{\Reynolds}
        \lapl \bu  \,, \quad \nabla \cdot \bu = 0
\,,
\ee{NavStokesDiff}
with boundary conditions $\bu = 0 $ at $y \pm 1$.  This equation is 
a little more complicated than \refeq{eqn:NavierStokes}, but having 
Dirichlet boundary conditions on $\bu$ makes the analysis much easier, 
since the set of allowable velocity fields (those fields that satisfy 
incompressibility and boundary conditions) forms a vector space. The equilibrium 
velocity fields we study start from $\bu$ which satisfies \refeq{NavStokesDiff}, and we may then add back the laminar part of the flow to produce physical fluid trajectories. 

\begin{figure}[!h]
 \begin{center} 
\includegraphics[width=0.5\textwidth]{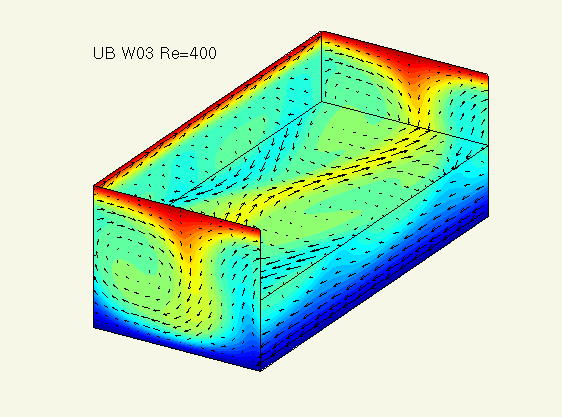}
  \caption{
Visualization of the upper branch equilibrium velocity field, from {\tt 
Channelflow.org}. 
   }
  \label{fig:UB}
  \end{center}
 \end{figure}

\subsection {\bf Computation of trajectories from equilibrium velocity fields}
\label{s:channelflow}

 In order to integrate streamlines of {\pCf}
and follow the paths of tracer particles, it is first
necessary to have numerically accurate \eqv\ $3D$-velocity fields.

The starting point for this task is to obtain the necessary data sets for 
evaluating velocity field values for a given \eqv, e.g. the upper branch 
as shown in \reffig{fig:UB}. These are made available at the website 
{\tt Channelflow.org} \cite{channelflow}. The data obtained 
\cite{channelflowDat} stores the spectral coefficients $\mathbf{\hat{u}}$ 
of the expansion of a velocity field $\mathbf{u(x)}$ satisfying \refeq{NavStokesDiff}. The form of the 
expansion is 
\begin{equation}
 \mathbf{u(x)} = \sum_{m_{y}=0}^{M_{y}-1}\sum_{m_{x}=0}^{M_{x}-1}\sum_{m_{z}=0}^{M_{z}-1}
 {\mathbf{\hat{u}}_{m_{x},m_{y},m_{z}} \bar{T}_{m_{y}}(y)e^{2\pi i(k_{x}x/L_{x} + k_{z}z/L_{z})}
 + \text{\small{(c.c.)}}}
\label{eqn:spectralsum}
 \end{equation}

The $\bar{T}(y)$'s are Chebyshev polynomials defined on the interval 
[a,b] (in most cases [-1,1]). For a given velocity field expansion, the 
upper bounds on the sums are known from the geometry, and the $k$'s are 
related to the $m$'s through the following relations: 
 \beq 
k_{x} = \left \{ 
\begin{array}{l}
m_{x} \hspace{20 mm} 0 \leq m_{x} \leq M_{x}/2   \\
m_{x} - M_{x} \hspace{10 mm} M_{x} < m_{x} < M_{x}  \\
\end{array}  \right.
\eeq 
\beq k_{z} = m_{z} \hspace{10 mm} 0 \leq m_{z} < M_{z}
\,.
\eeq
Hence, with a knowledge of the spectral coefficients we can compute 
$\mathbf{u(x)}$ by evaluating this sum at a particular $\bx = (x,y,z)$. 

Various internal functions within {\tt Channelflow.org} have been written 
to compute $\bu$ on a set of gridpoints. It is possible, by interpolation 
of the velocity fields on these gridpoint values, to integrate a 
trajectory with great computational speed. However this will not be 
nearly as accurate as evaluating the sum \refeq{eqn:spectralsum} 
directly. So we evaluate \refeq{eqn:spectralsum} to give the exact 
velocity field at every point along a trajectory, adding back the laminar part of the flow. We are able to perform 
these computations in Matlab with enough speed to compute many tracer 
particle trajectories within an equilibrium velocity for an adequate 
length of time to study the flow dynamics.  The code has been checked to 
be correct by picking an $(x,y,z)$ coordinate that \emph{happens} to lie 
on a gridpoint value and then comparing the result to the value given by 
the internal {\tt Channelflow.org} functions.

\subsection{Symmetries of {\pCf}}
\label{s:PCF_symm}

As part of our theoretical analysis of trajectories of fluid particles within an equilibrium velocity field, it will be critical to use and understand the symmetries involved in the special geometry of {\pCf}. Thus we take a quick detour to discuss these symmetries from a group-theoretic perspective. We focus on the symmetries relevant to the equilibria studied in this work; additional details are provided in \cite{HalcrowThesis}.

\PCf\ is invariant under two reflections $\sigma_1,\sigma_2$ and a
continuous two-parameter group of translations $\tau(\shift_x, \shift_z)$:
\begin{align}
\sigma_1 \, [u,v,w](x,y,z) &= [u, v,-w](x,y,-z) \nnu \\
\sigma_2 \, [u,v,w](x,y,z) &= [-u,-v,w](-x,-y,z)  \label{reflSfit1}\\
\tau(\shift_x, \shift_z)[u,v,w](x,y,z) &= [u,v,w](x+\shift_x,y,z+\shift_z) \nnu\,.
\end{align}
The \NSe s and boundary conditions are invariant for any symmetry $s$
in the group generated by these elements:
$\partial (s \bu) / \partial t = s (\partial \bu / \partial t)$.

The plane Couette symmetries can be interpreted geometrically in the space of
fluid velocity fields. Let $\bbU$ be the space of
square-integrable, real-valued velocity fields that satisfy the kinematic
conditions of \pCf:
\begin{align}
 \bbU  &= \{\bu \in L^2(\Omega) \; | \; \grad \cdot \bu = 0,
               \; \bu(x, \pm 1, z) = 0, \notag  \\
         &\phantom{=} {} \qquad \qquad \qquad \; \; 
          \bu(x, y, z) = \bu(x+L_x, y, z) = \bu(x, y, z + L_z)\}  \,.
\end{align} 
The continuous symmetry $\tau(\shift_x, \shift_z)$ maps each state
$\bu \in \bbU$ to a $2D$ torus of states with identical dynamic
behavior. This torus in turn is mapped to four equivalent tori by
the subgroup $\{1,\sigma_1,\sigma_2, \sigma_1 \sigma_2\}$. In
general a given state in $\bbU$ has four $2D$ tori of dynamically
equivalent states.

Most of the Eulerian \eqva\ that are currently known for \pCf\
are invariant under the `shift-reflect' symmetry
$s_1 = \tau(L_x/2,0) \, \sigma_1$ and the `shift-rotate' symmetry
$s_2 = \tau(L_x/2,L_z/2) \, \sigma_2$.  These symmetries form a group
\beq
S = \{1, s_1, s_2, s_3\}, \qquad s_3 = s_1 s_2, 
\eeq
which is isomorphic to the Abelian dihedral group $D_2$, and is a 
subgroup of a larger group generated by plane Couette symmetries. The 
group acts on velocity fields as: 
\begin{align}
s_1 \, [u, v, w](x,y,z) &= [u, v, -w](x+L_x/2,\, y,\, -z) \nnu \\ 
s_2 \, [u, v, w](x,y,z) &= [-u, -v, w](-x+L_x/2,\,-y,\,z+L_z/2) \label{shiftRot} \\
s_3 \, [u, v, w](x,y,z) &= [-u,-v,-w](-x,\, -y,\, -z+L_z/2)  \nnu 
\,
\end{align}

We denote the $S$-invariant subspace of states invariant under
symmetries \refeq{shiftRot} by
\begin{align}
\bbUsymm  &= \{\bu \in \bbU  \: | \;
              s_j \bu = \bu\,, \;\;  s_j \in S \}
\,,
\label{symmSubspU}
\end{align}

where $ \bbUsymm \subset \bbU$.
$\bbUsymm$ is a flow-invariant subspaces: states initiated
in it remain there under the \NS\ dynamics.

Translations of half the cell length in the spanwise and/or streamwise
directions commute with $S$. These operators generate a discrete
subgroup of the continuous translational symmetry group $SO(2) \times
SO(2)$ :
\beq
T = \{e,\tau_x,\tau_z,\tau_{xz}\}
    \,,\qquad
    \tau_x = \tau(L_x/2,0)
    \,,\;
    \tau_z = \tau(0,L_z/2)
    \,,\;
    \tau_{xz} = \tau_x \tau_z
\,.
\ee{tauD2}
Since the action of $T$ commutes with that of $S$,
the three half-cell translations $\tau_x \bu, \, \tau_z \bu,$ and
$\tau_{xz} \bu$ of $\bu \in \bbUsymm$ are also in $\bbUsymm$.

We know that the equilibria  $EQ_1$-$EQ_8$ are symmetric in $S$ because 
they satisfy those symmetries numerically. There is no a priori reason 
that the equilibria should be $S$-symmetric, other than $S$ symmetry 
fixes $x,z$ phase and so rules out relative equilibria. But $s_3$ 
symmetry alone does the same, and a few equilibria are known that have 
$s_3$ symmetry but neither $s_1$ nor $s_2$ symmetry. There are equilibria 
with other symmetries that fix $x,z$ phase but have other translations 
than the half-cell shifts. 

It is also possible to form other isotropy subgroups from the plane 
Couette symmetries $\tau_x$, $\tau_z$, $\sigma_1$, $\sigma_2$. These 
elements generate a group $G$ of order 16, of which there are various 
subgroups of possible orders $\{1,2,4,8,16\}$. It is known that other 
equilibria posses different symmetries, corresponding to different 
subgroups of $G$. For example, for equilibrium $EQ_8$, we find there is 
symmetry under an invariance group of order 8, denoted $S_8$, that is 
isomorphic to the dihedral group $D_4$. 
\[
S_8 = \{e, s1, s2, s3, s4, s5, s6, s7\}
\]
where $s_4 = \tau_z \, \sigma_1$, $s_5 = s_4 s_2$, $s_6 = \tau_x \tau_z$, $s_7 = \sigma_2$. The action of these additional symmetries of $S_8$ on velocity fields is:
\begin{align}
s_4 \, [u, v, w](x,y,z) &= [u, v, -w](x,\, y,\, -z + L_z/2) \nnu \\ 
s_5 \, [u, v, w](x,y,z) &= [-u, -v, -w](-x+L_x/2,\,-y,\,-z) \label{S_8} \\
s_6 \, [u, v, w](x,y,z) &= [u,v,w](x+L_x/2,\, y,\, z+L_z/2)  \nnu  \\
s_7 \, [u, v, w](x,y,z) &= [-u,-v,w](-x,\, -y,\, z)  \nnu 
\,
\end{align}

Which symmetries happen to exist for the different equilibria will have 
important implications for studying the dynamics of the flow. 

\subsection{Symmetry and {\stagp}s}
\label{s:symm_stag}

From the form of $s_3$ in \refeq{shiftRot}, we can see that any Eulerian equilibrium that
is invariant under $S$ has 4 Lagrangian \stagp s at which the velocity is 0,
which satisfy the condition:
\begin{equation}
 (x,y,z) = (-x, -y, -z+L_z / 2) \label{shiftRot_eqva}
\end{equation}
There are 4 points which satisfy this constraint:
\bea
  \bold{x}_{_{SP_{1}}} &=& (L_x/2,0,L_z/4) \continue
  \bold{x}_{_{SP_{2}}} &=& (L_x/2,0,3L_z/4) \continue
  \bold{x}_{_{SP_{3}}} &=& (0,0,L_z/4) \label{s3lagrange} \\
  \bold{x}_{_{SP_{4}}} &=& (0,0,3L_z/4) \nnu
 \,.
\eea

We refer to these as {\stagp}s $SP_1$-$SP_4$. Due to the periodic 
boundary conditions, we equivalently have 
 $(L_x,0,L_z/4)=SP_3$ and $(L_x,0,3L_z/4)=SP_4$.
Also of note is the fact that there can exist no $s_3$-invariant \reqva, 
since $s_3$ operation flips both the $x$ and $z$ axes. These {\stagp}s 
will exist in all of the equilibria with $S$-symmetry. Additionally, for 
an equilibrium such as $EQ_8$ which possesses $S_8$ symmetry, from the 
action of $s_5$ in \refeq{S_8}, we will find {\stagp}s wherever 
\beq
 (x,y,z) = (-x+L_x/2, -y, -z) 
 \,,
\ee{second_condition}
which gives the additional points:
\bea
  \bold{x}_{_{SP_{5}}} &=& (L_x/4,0,0) \continue
  \bold{x}_{_{SP_{6}}} &=& (3L_x/4,0,0) \continue
  \bold{x}_{_{SP_{7}}} &=& (L_x/4,0,L_z/2) \label{s3lagrange} \\
  \bold{x}_{_{SP_{8}}} &=& (3L_x/4,0,L_z/2) \nnu
 \,.
\eea

In fact, we can generalize the discussion. Looking at the way the plane 
Couette symmetries act on velocity fields in \refeq{reflSfit1}, we see 
that since $\tau$ does not affect the velocity components, the condition 
needed to produce a {\stagp} (in which all three velocity components are 
negated at some shifted position) will work only for the combinations of 
these elements which contain both $\sigma_{1}$ and $\sigma_{2}$ an odd 
number of times. Within the group $G$ of order 16 of plane Couette 
symmetries generated by $\sigma_{1}$, $\sigma_{2}$, $\tau_{x}$, 
$\tau_{z}$, the requirement means we just have to identify elements that 
have a $\sigma_{1}\sigma_{2}$ term. 

There are in fact four such elements of $G$ that contain a
$\sigma_{1}\sigma_{2}$ term. We denote these as $g_1 = \sigma_{1}\sigma_{2}$,
$g_2 = \sigma_{1}\sigma_{2}\tau_{x}$, $g_3 =
\sigma_{1}\sigma_{2}\tau_{z}$, and $g_4 = \sigma_{1}\sigma_{2}\tau_x
\tau_z$. 
\begin{align}
g_1 \, [u,v,w](x,y,z) &= [-u,-v,-w](-x,-y,-z)  \\
g_2 \, [u,v,w](x,y,z) &= [-u,-v,-w](-x+L_{x}/2,-y,-z)  \\
g_3 \, [u,v,w](x,y,z) &= [-u,-v,-w](-x,-y,-z+L_{z}/2)  \\
g_4 \, [u,v,w](x,y,z) &= [-u,-v,-w](-x+L_{x}/2,-y,-z+L_{z}/2)
\end{align}

Different isotropy subgroups of $G$ may or may not contain a symmetry 
which corresponds to one of these $g_1$-$g_4$ elements, however any $g_i$ 
that is part of an invariance group for an equilibrium implies the 
existence of four symmetrically-located \stagp s in the $y = 0$ plane. 
Note that $g_3$ and $g_2$ are the elements already  discussed that 
produce $SP_1$-$SP_8$. 

\begin{figure}[!h]
\includegraphics[width=1.0\textwidth]{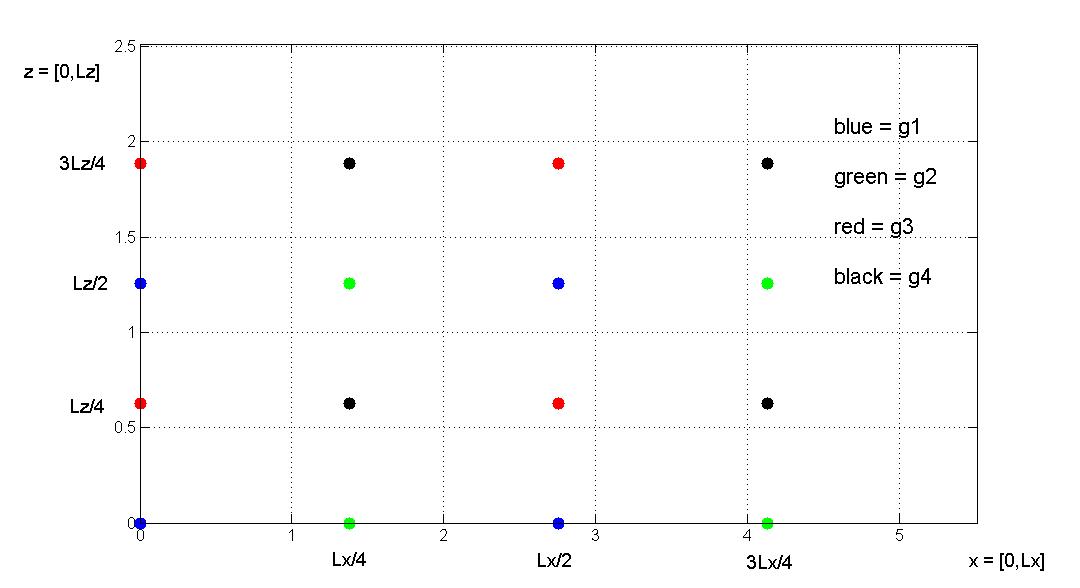}
  \caption{
   Sets of possible \stagp s. If one of the $g_i$ symmetries is
   possessed, the velocity field will have \stagp s of the color
   corresponding to that symmetry.
   }
  \label{fig:stags7_26}
 \end{figure}

Any equilibrium with $g_1$ symmetry implies that there would additionally 
be \stagp s at $(0,0,0)$, $(L_{x}/2,0,0)$, $(0,0,L_{z}/2)$, and 
$(L_{x}/2,0,L_{z}/2)$. And similarly, $g_4$ symmetry implies the 
existence of \stagp s at $(L_{x}/4,0,L_{z}/4)$, $(L_{x}/4,0,3L_{z}/4)$, 
$(3L_{x}/4,0,L_{z}/4)$, and $(3L_{x}/4,0,3L_{z}/4)$. The set of all 
possible {\stagp}s based on various \pCf\ symmetries is shown in 
\reffig{fig:stags7_26}. 

So the question of existence of \stagp s for a given equilibrium is, 
which of the $g_i$ symmetries does that equilibrium possess? This is a 
question related to invariance under the isotropy subgroups. Of 
importance, this does not address the question of whether \textit{other} 
nontrivial \stagp s may exist that are not based on symmetry arguments 
alone. For the known equilibria of {\pCf} $EQ_1$-$EQ_{11}$, all of them 
have $g_3$ symmetry and $EQ_7$, $EQ_8$ additionally have $g_2$ symmetry. 
This is likely related to the fact that searches for equilibria were done 
in a symmetric subspace which contained the $g_3$ elements (the 
$S$-symmetric subspace). 

\subsection{Any nontrivial \stagp\ has a partner, symmetric about another {\stagp}}

Though our symmetry arguments do not determine whether or not there may exist \textit{additional} {\stagp}s which are not forced by the $g_i$ symmetries in the preceding section, we can in fact that show that for equilibria which exist in one of the flow-invariant subspaces that contains a $g_i$-symmetry (for example, $S$ has $g_3$ symmetry and $S_8$ has both $g_2$ and $g_3$ symmetry), any additional nontrivial {\stagp}s that exist must occur in symmetric pairs centered around the other known {\stagp}s.

Consider one of the equilibria in the $S$-invariant subspace, such as $EQ_2$. Again, the
 action of $s_3 \in S$ on velocity fields gives:
 \beq    s_3 \, [u, v, w](x,y,z) = [-u,-v,-w](-x,\, -y,\, -z+L_z/2)\nnu\, .
 \eeq
 If $(x_{_{SP}},y_{_{SP}},z_{_{SP}})$ is a \stagp, $[u, v,
 w](x_{_{SP}},y_{_{SP}},z_{_{SP}}) = [0,0,0]$, then
 \begin{align} s_3 \, [u, v, w](x_{_{SP}},y_{_{SP}},z_{_{SP}}) &= [-u,-v,-w](-x_{_{SP}},\, -y_{_{SP}},\, -z_{_{SP}}+L_z/2) \nnu\, \\
 &= [0,0,0](-x_{_{SP}},\, -y_{_{SP}},\, -z_{_{SP}}+L_z/2) .
 \end{align}
 Thus $(-x_{_{SP}},\, -y_{_{SP}},\, -z_{_{SP}}+L_z/2)$ is also a \stagp.

We may parameterize a line passing through two points 
$(x_{1}, y_{1}, z_{1}),(x_{2}, y_{2}, z_{2})$
 as
 \begin{align}
  x &= x_{1} + (x_{2} - x_{1})t \\
  y &= y_{1} + (y_{2} - y_{1})t \\
  z &= z_{1} + (z_{2} - z_{1})t \\
  t &\in (-\infty,\infty) .
 \end{align}

 Using the two stagnation points $(x_{_{SP}},y_{_{SP}},z_{_{SP}})$ and $(-x_{_{SP}},-y_{_{SP}},-z_{_{SP}} + L_z/2)$ this becomes
 
 \begin{align}
  x &= x_{_{SP}}(1-2t) \\
  y &= y_{_{SP}}(1-2t) \\
  z &= z_{_{SP}}(1-2t) + \frac{L_{z}}{2} t .
 \end{align}
When $t = 1/2$ this system returns $(x,y,z) = (0,0,L_{z}/4)$, showing 
that $SP_3$ lies on the line between these two \stagp s, halfway in 
between them. 

If we invoke the box periodicities: $x = x + L_{x}$, $z = z + L_{z}$, it 
is easy to show that this pair of {\stagp}s is also symmetric about any 
of $SP_1$-$SP_4$. For example, \\ 

 \noindent$\mathbf{x = x + L_{x}}$:

 \noindent $(x_{_{SP}},y_{_{SP}},z_{_{SP}})$ is a \stagp\ $\Rightarrow$
 $(-x_{_{SP}}+L_{x},-y_{_{SP}},z_{_{SP}}+L_{z}/2)$ a \stagp.
 \begin{align}
  x &= x_{_{SP}}(1-2t) + L_{x}t \\
  y &= y_{_{SP}}(1-2t) \\
  z &= z_{_{SP}}(1-2t) + \frac{L_{z}}{2} t .
 \end{align}
When $t = 1/2$ this returns $(x,y,z) = (L_{x}/2,0,L_{z}/4)$, so that the 
new stagnation points lie symmetrically on a line passing through $SP_1$. 

For an equilibrium invariant under $S_8$, such as $EQ_8$, existence of 
any additional nontrivial {\stagp} will then imply \textit{two} 
additional {\stagp}s, based on the action of $g_2$ and $g_3$. 
 If $(x_{_{SP}},y_{_{SP}},z_{_{SP}})$ is a \stagp, then  
 $(-x_{_{SP}},\, -y_{_{SP}},\, -z_{_{SP}}+L_z/2)$ and 
 $(-x_{_{SP}} + L_x/2,\, -y_{_{SP}},\, -z_{_{SP}})$ are also \stagp s. 

We will investigate numerical methods to determine the possible existence 
of any nontrivial {\stagp}s. In fact for $EQ_2$, as we show in the next 
section, we do find such a point and it's symmetric partner. These 
additional {\stagp}s are critical for understanding the flow dynamics in 
the equilibrium field, as their stable and unstable manifolds provide us 
with an outline of the overall dynamics. 

\section{Lagrangian dynamics}
\label{s:Lagrangian}

We know of the existence of  \stagp s in the flow of an equilibrium 
velocity field predicted from the symmetries of \pCf. Thus the starting 
point for our investigation is clear; treating an equilibrium velocity 
field as an autonomous dynamical system we have already identified the 
"fixed points" of the system, which we refer to in this context as the 
\stagp s.  Using the sum formula for computing velocities at any point in 
the \pCf\ domain \refeq{eqn:spectralsum}, by differentiating this formula 
it is a simple matter to compute the $[3\!\times\! 3]$ velocity gradients 
or Jacobian matrix at any point. Eigenvalues and eigenvectors of this 
matrix will provide linear stability analysis results for the {\stagp}s, 
and allow us to compute and visualize the local stable and unstable 
manifolds by starting a collection of tracer points along the directions 
of the eigenvectors, integrating them forwards and backwards in time 
(when the local tangent space is $2D$, trajectories are started 
throughout a small radius in the plane spanned by the eigenvectors). 
Though this method may underrepresent a part of the manifold for the $2D$ 
case \cite{SahVla09}, we find that the approximation works for revealing 
the interesting and relevant dynamical behaviors we seek. 

In order to investigate additional locations in the domain for which no 
movement occurs, we may numerically compute $|\bu|^{2}$ along a fine grid 
and try to ascertain regions where the velocity value falls below a given 
threshold. Then, using interpolation within these regions, any additional  
\stagp s can be pinned down. 

With the determination of the {\stagp}s and their invariant manifolds, we 
find a natural way to view the physical space of the fluid, partitioned 
into regions wherein the dynamics is dominated by the trajectories 
following closely to the manifolds themselves. This provides us with a 
framework for studying how transport may occur within and between the 
different regions. 

\subsection{The upper branch equilibrium}
\label{s:eq2}

Our analysis is carried out for the upper branch \eqv\ velocity field, 
$EQ_2$, at $\Reynolds = 400$. 
The cell size parameters are 
\beq   
[L_x,2,L_z]
         = \; [2\pi/1.14,2,4\pi/5]
         ~ [5.512,2,2.513].
\ee{cellW03}

\begin{figure}[!h]
\centering
    \begin{subfigure}{0.98\textwidth}
    \includegraphics[width=1.0\textwidth]{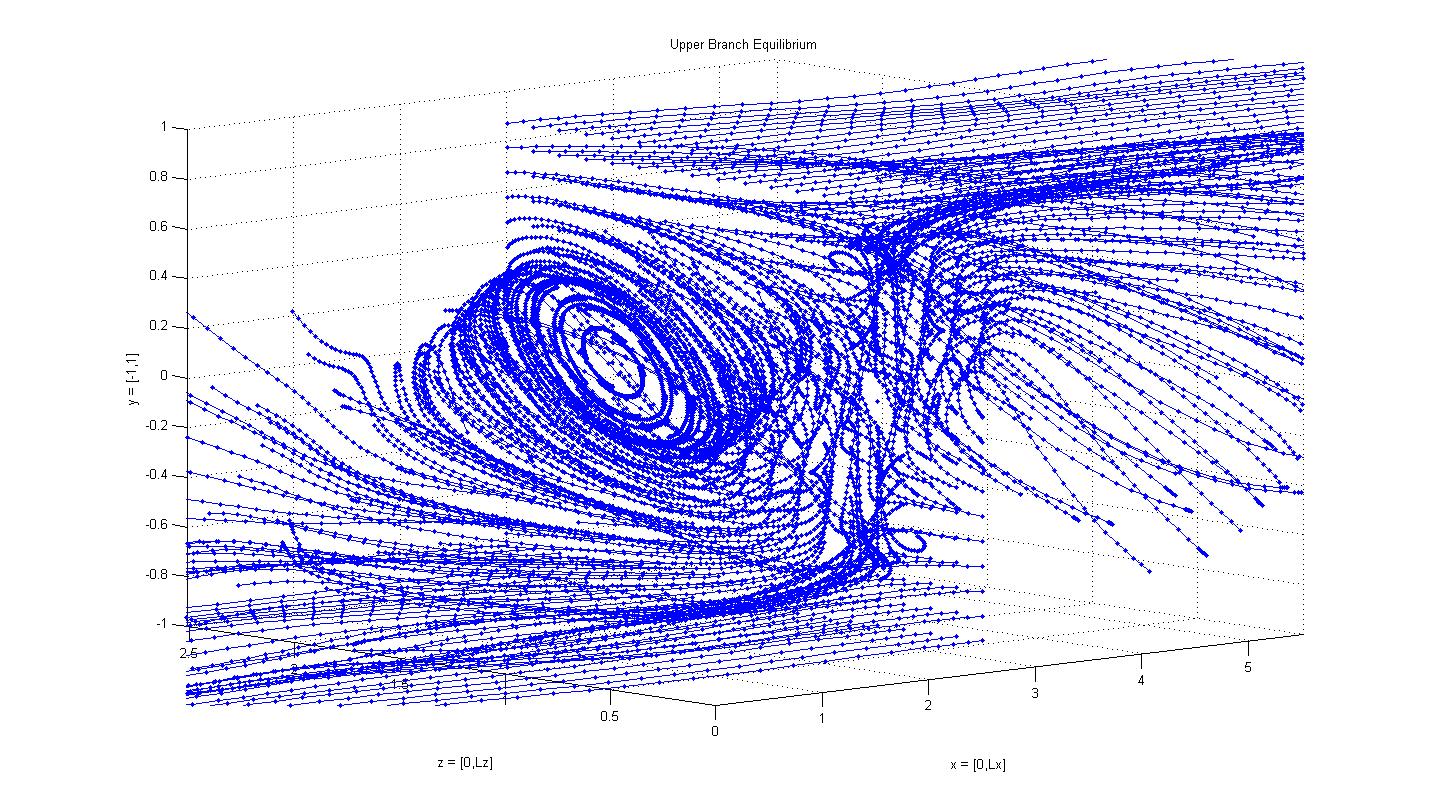}
      \caption{
        $3D$ perspective view
       }
      \label{fig:UBs_a}
    \end{subfigure}

    \begin{subfigure}{0.98\textwidth}
    \includegraphics[width=1.0\textwidth]{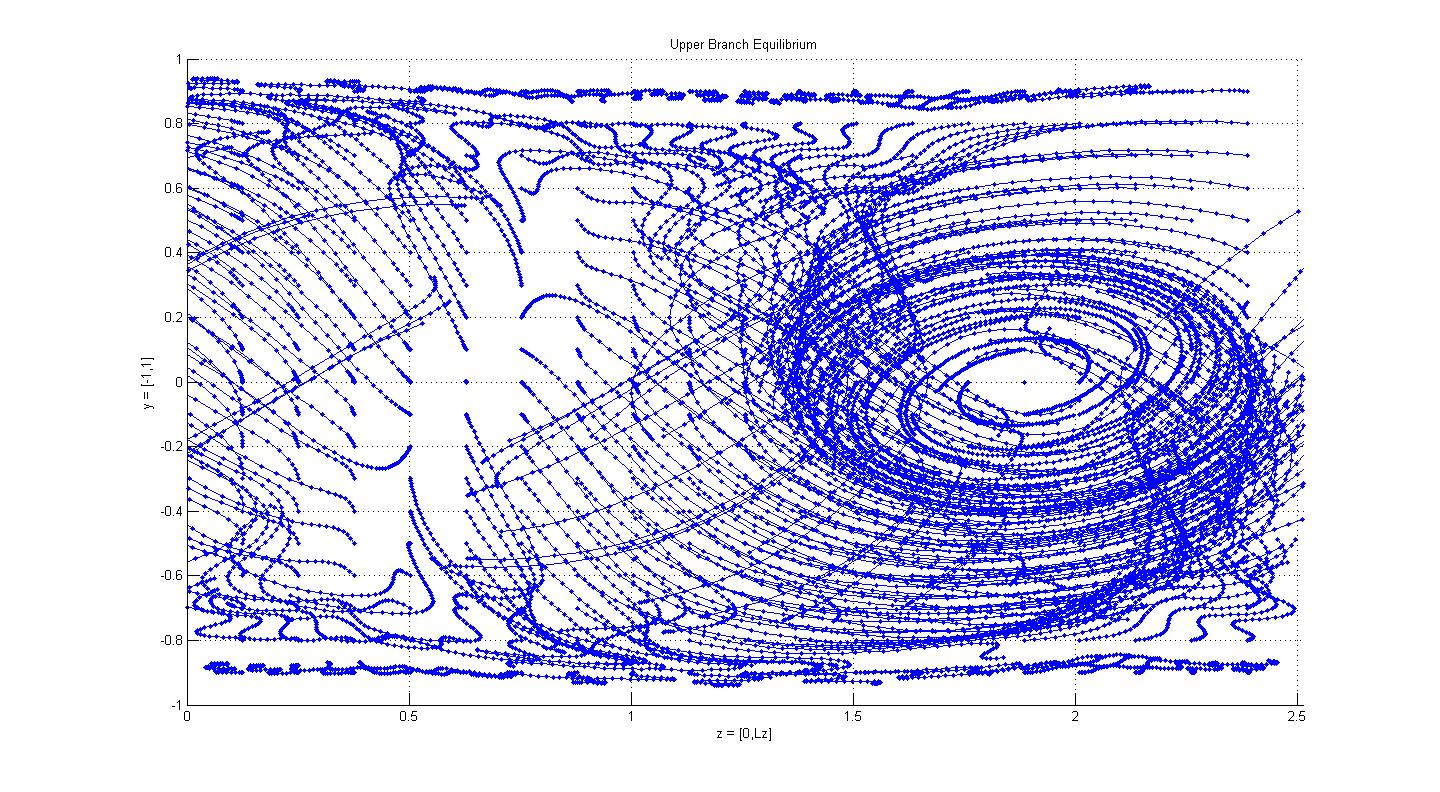}
      \caption{
        Rotated to show the 2 \stagp s
       }
      \label{fig:UBs_b}
    \end{subfigure}
    \caption{
Grid of $19 \times 19$  initial points in the $[y,z]$ plane, centered at 
$x = L_x/2$; integrated for 15 time units to produce tracer particle 
trajectories for $EQ_2$.} 
\label{fig:UBs}
 \end{figure}

To begin, we look at the evolution of Lagrangian tracers starting on a 
grid of points, shown in \reffig{fig:UBs}. The grid is chosen to lie 
in the $[y,z]$ plane, centered at $x = L_x/2$. The initial points are 
equally spaced, and offset by one position from the edge of the box. If 
the number of points is chosen to be one less than a multiple of 4, there 
will be points starting at $\bold{x}_{_{SP_{1}}}=(L_x/2,0,L_z/4)$ and 
$\bold{x}_{_{SP_{2}}}=(L_x/2,0,3L_z/4)$. The trajectories are integrated 
and run for a relatively short time. Just from evolving the 
grid of points alone, we begin to get a feel for the dynamics and start 
to see the formation of interesting patterns and vortical structures. 

$EQ_2$ invariance under the symmetry group $S$, explained  in 
\refsect{s:symm_stag}, implies the existence of 4 \stagp s 
$SP_1$-$SP_4$, \refeq{s3lagrange}. In \reffig{fig:UBs_b} the view 
from \reffig{fig:UBs_a} has been rotated in order to reveal two of 
these \stagp s. The visualization of the behavior of trajectories near 
these fixed points reveals their  qualitative nature. The point at 
$3L_z/4$ in \reffig{fig:UBs_b} appears to be an unstable spiral, 
whereas the point at $L_z/4$ is hyperbolic. In order to verify these 
hypotheses, eigenvalues and stable/unstable manifolds for these \stagp s 
are computed. 

\subsection{Linearization and stability}

For a perturbation $\delta$\bx\ away from one of the {\stagp}s,
the change in the velocity field is given by $\delta\bu = \Mvar
\delta\bx$ where $\Mvar$ is the nine component \velgradmat\ defined
by $\Mvar_{ij}=\frac{\partial u_{i}}{\partial x_{j}}$. Since \bu\ is
given by \refeq{eqn:spectralsum}, it is a relatively simple
extension of this formula to evaluate these partials. To find
$\partial\bu/\partial y$, one needs to use the relation
$\frac{\partial}{\partial y}T_{n}(y) = n U_{n-1}(y)$ where $T_{n}$
is the $n$th Chebyshev polynomial of the first kind and $U_{n}$ is
the $n$th Chebyshev polynomial of the second kind. Everything else
is straightforward.
The eigenvalues of $\Mvar$, evaluated at a {\stagp}, determine local stability
and reveal the qualitative nature of the motion nearby the \stagp.
For the \stagp s $SP_1$ - $SP_4$, the eigenvalues, eigenvectors,
and velocity gradients matrices are as follows. \\

$\bold{x}_{_{SP_{1}}}=(L_x/2,0,L_z/4)$: There are 3 real eigenvalues, two 
positive and one negative. 
\begin{align}
&\eigExp[1] = -0.4652099 \,,\quad
\jEigvec[1] =
\begin{pmatrix}
             {0.9844417} \cr
             {0.1743315} \cr
             {0.0219779} \cr
   \end{pmatrix} \\ \label{sp1_evec1}
    &\eigExp[2] = 0.4008961 \,,\quad \jEigvec[2] =
\begin{pmatrix}
             {0.5704000} \cr
             {-0.7666749} \cr
             {0.2947091} \cr
   \end{pmatrix} \\  \label{sp1_evec2}
    &\eigExp[3] = 0.0643139 \,,\quad \jEigvec[3] =
\begin{pmatrix}
             {0.4082166} \cr
             {0.7525949} \cr
             {0.5166819} \cr
   \end{pmatrix} \\ \label{sp1_evec3} 
   \end{align}
   The \velgradmat\ is
\beq
   \Mvar =
   \begin{pmatrix}
   {-0.4305385} &  {-0.3002042} &{0.8282447} \cr
   {-0.1221356} &   {0.2456107} & {-0.1675796} \cr
   {0.0001651}  &   {-0.0828951}  & {0.1849278} \cr
            \end{pmatrix}
\eeq
The point is a saddle; it has 1 stable dimension and a $2D$ plane of 
instability spanned by $\jEigvec[2]$ and $\jEigvec[3]$, with the 
eigenvalues summing to 0, as required by a volume-preserving flow. 
    
The \stagp\ $SP_4$ at $(0,0,3L_z/4)$ has the same eigenvalues as for 
$SP_1$. It's eigenvectors and \velgradmat\ differ by a minus sign in the 
third component (except for $\Mvar_{33}$ where the two minuses cancel). 

$\bold{x}_{_{SP_{2}}}=(L_x/2,0,3L_z/4)$: 
There is one real, negative eigenvalue and a complex
pair with positive real part.

\begin{align}
&\eigExp[1] = -0.0352362 \,,\quad \jEigvec[1] =
\begin{pmatrix}
             {-0.9452459} \cr
             {-0.1893368} \cr
             {-0.2658228} \cr
   \end{pmatrix}
   \\
&\eigRe[2] \pm i\,\eigIm[2] = 0.0176181 \pm i\,0.0862176
   \\
&\jEigvec[2] =
\begin{pmatrix}
             {0.3737950 + 0.0544113i} \cr
             {0.2098940 - 0.4925773i} \cr
             {0.7554000} \cr
   \end{pmatrix}
\,,\quad
\jEigvec[3] =
\begin{pmatrix}
             {0.3737950 - 0.0544113i} \cr
             {0.2098940 + 0.4925773i} \cr
             {0.7554000} \cr
   \end{pmatrix}
\nnu\,.
\end{align}
The \velgradmat\ is 
\[ 
   \Mvar =
   \begin{pmatrix}
   {-0.0316935} & {-0.0708737} &  {0.0378835} \cr
  {-0.0250579} & {-0.0218884} &  {0.0795969} \cr
   {0.0014742} & {-0.1320575} &  {0.0535818} \cr
   \end{pmatrix}
\] 
Trajectories starting near this \stagp\ spiral out in a plane spanned by 
the complex pair of eigenvectors. The stable direction is one-dimensional 
and points primarily along the $x$ direction. 
    
$SP_3$ at $(0,0,L_z/4)$ has the same eigenvalues as $SP_2$ and again, the 
\velgradmat\ is the same except for sign changes in the third component. 
This follows from the plane Couette symmetries. 

\subsection{Further {\stagp}s}

Having analyzed {\stagp}s $SP_1$-$SP_4$, before further investigating the 
dynamics, it is natural to wonder whether other such {\stagp}s may exist 
that do not necessarily follow from a symmetry argument. To answer this 
question, as mentioned above, we numerically compute $|\bu|^{2}$ along a 
fine grid and look for where it's value falls below a given threshold. 

We create a more refined grid of velocities which is $144 \times 105 
\times 144$. This is three times the 48 $\times$ 35 $\times$ 48 grid in 
each dimension used to show the initial tracer trajectories, and contains 
about 2.2 million points. At each point $|\bu|^{2}$ is then calculated 
and at every point that satisfies $|\bu|^{2} < \epsilon$ for some 
arbitrarily chosen $\epsilon$, the point is plotted. 

\begin{center}
\begin{figure}[!h]
\includegraphics[width=1.0\textwidth]{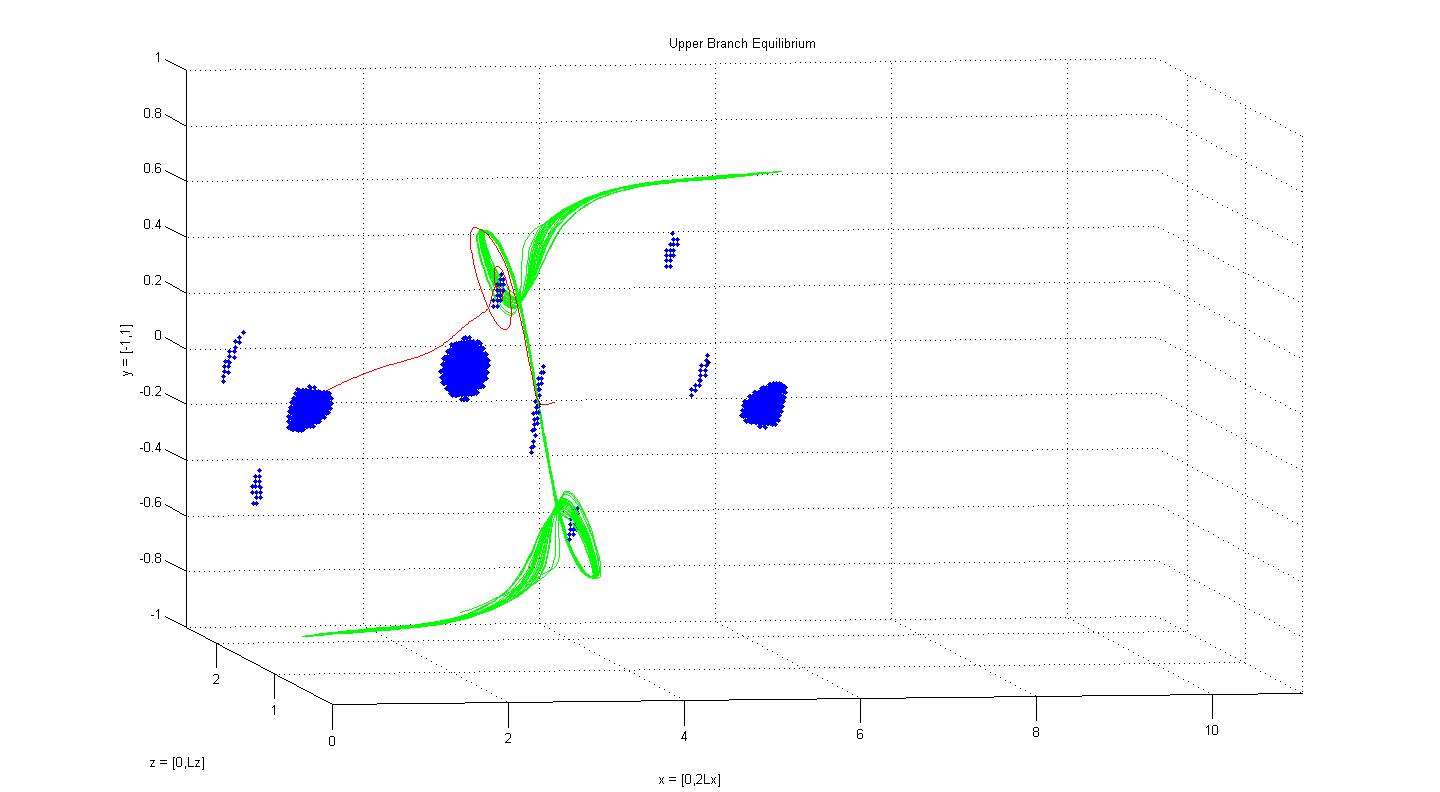}
  \caption{
Blue clumps of points indicate where the velocity for $EQ_2$ is very 
close to zero. Shown along with the stable manifold of $SP_3$ and the 
unstable manifold of $SP_1$. 
          }
  \label{fig:fine_usquare}
 \end{figure}
\end{center}

In \reffig{fig:fine_usquare} we show regions in the cell where 
$|\bu|^{2}$ is very small for $\epsilon = 10^{-4}$, notated by the globs 
of blue dots. The trajectories shown along with the points of small 
velocity in this figure, explained below, are also suggestive of the 
existence of a {\stagp} within the spiraling region. The four previously 
known {\stagp}s are identified in the figure, but we also see a couple of 
additional clumps. Honing in one of the suspicious clusters, starting 
from the gridpoint value with smallest velocity in the suspicious region, 
$\bx_{0} \approx (2.33476, 0.40952, 0.64577)$, and its reflection through 
$\bold{x}_{_{SP_{1}}}$, $\bx_{0}' =2 \bold{x}_{_{SP_{1}}} - \bx_{0}$, the 
Newton iteration 
\[ 
 \bx_{k+1} = \bx_{k} -
          {\Mvar}^{-1}(\bx_{k}) \, \bu(\bx_{k})
\] 
converges rapidly to verify \textit{another} pair of \stagp s. Because we 
have already used notation to define points $SP_1$-$SP_8$ in 
\refsect{s:symm_stag}, we refer to these new numerically discovered 
{\stagp}s as $SP_{N1}$ and $SP_{N2}$: 

\begin{align}
&\bold{x}_{_{SP_{N1}}} =(2.35105561774981,0.42293662349708,0.65200166068573)
\\
&\bold{x}_{_{SP_{N2}}}=(3.16051044117966,-0.42293662349708,0.60463540075018)
\label{eqn:newspNewt}
\,.
\end{align}
We see the
 symmetry in the $y$-component of this pair, and in fact
these points are shown to be
 symmetric about the point $SP_1$, as discussed in \refsect{s:symm_stag}:
 \beq
    (\bold{x}_{_{SP_{N1}}} +\bold{x}_{_{SP_{N2}}})/2 = \bold{x}_{_{SP_{1}}}
 \,.
 \eeq

Repeating the linear stability analysis for $SP_{N1}$ and $SP_{N2}$: 
There is one real, positive eigenvalue and a complex pair with negative 
real part. 
  \begin{align} &\eigExp[1] = 0.1453207 \,,\quad \jEigvec[1] =
\begin{pmatrix}
             {0.9307982} \cr
             {0.3502306} \cr
             {0.1046576} \cr
   \end{pmatrix}
   \\
&\{ \eigExp[2],\eigExp[3]\}
  = \eigRe[2] \pm i \,\eigIm[2] =  -0.0726603 \pm i\, 0.3733478
   \nnu\\
&\jEigvec[2] =
\begin{pmatrix}
             {~0.5226203} \cr
             {-0.6703938} \cr
             {~0.2065610} \cr
   \end{pmatrix}
    \,,\quad
\jEigvec[3] =
\begin{pmatrix}
             {~0.3779843} \cr
             {~
             0} \cr
             {- 0.3031510} \cr
   \end{pmatrix}
\,.
\end{align}
The \velgradmat\ is
\[ 
   {\Mvar} =
   \begin{pmatrix}
   {0.0225166} &  {0.0985763} &{0.7623083} \cr
   {0.1714566} &   {-0.1275193} & {-0.6118476} \cr
   {-0.0615378}  &   {0.1755954}  & {0.1050028} \cr
            \end{pmatrix}
\,.
\] 

We have this time a $1D$ unstable manifold and a $2D$ spiraling stable 
manifold. The trajectories shown in \reffig{fig:fine_usquare}, which 
originate close to $SP_1$ and $SP3$, wander close to the spiraling stable 
manifold of the numerically discovered $SP_{N1}$, showing how the 
dynamics tends to be dominated by these {\stagp}s. 

 \begin{figure}[!h]
\includegraphics[width=1.0\textwidth]{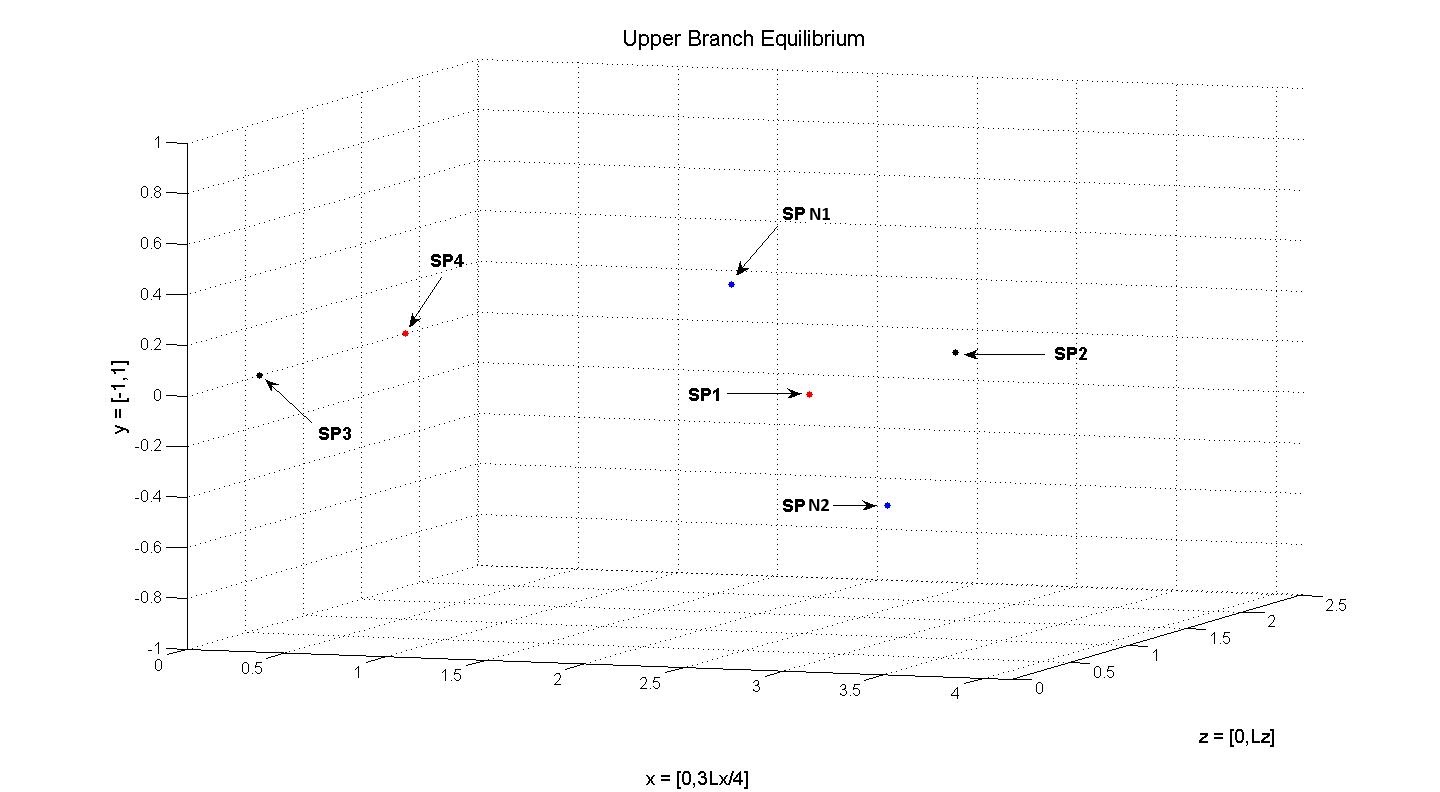}
  \caption{
   The 6 unique \stagp s within one periodic box for $EQ_2$. 
   $SP_1$-$SP_4$ are guaranteed by $EQ_2$ symmetries, $SP_{N1}$ and 
   $SP_{N2}$ are determined numerically. 
   }
  \label{fig:stagps_label}
 \end{figure}

 \begin{figure}[!h]
\includegraphics[width=1.0\textwidth]{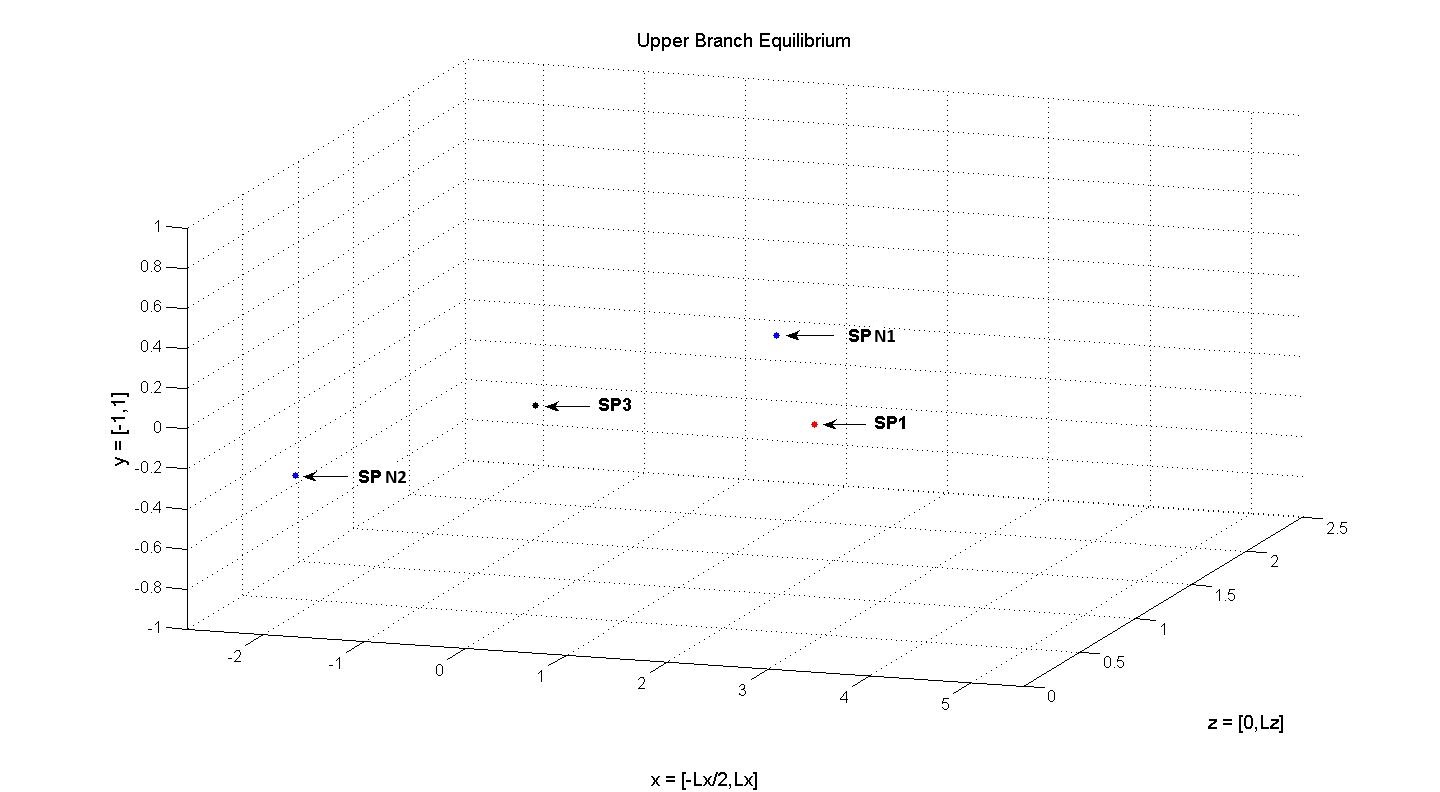}
  \caption{
   The 4 \stagp s that occur within the domain $\Omega$.
   }
  \label{fig:stagps_label2}
 \end{figure}

We have been describing all \stagp s which are inside a single periodic 
cell with dimensions $L_x \times 2 \times L_z$, pictured in 
\reffig{fig:stagps_label}. However even within this cell there is a 
redundancy in labeling all of these points as distinct. The interesting 
dynamics and connections between the different \stagp s occur along the 
$x$ direction. To understand what is happening one needs to look only at 
a subset of these \stagp s that lies in the right or left half of the 
box, that is, in the interval $[0,L_{z}/2]$ or the interval 
$[L_{z}/2,L_{z}]$. We have chosen the interval $[0,L_{z}/2]$. In the $x$ 
direction the most convenient interval is not actually $[0,L_{x}]$, 
rather we look at the \stagp s in the open interval $(-L_{x}/2,L_{x})$, 
open so as to ignore the repeated translations on the boundary. Thus an 
alternate domain of investigation that will be convenient to sometimes 
use is 
\[ 
\Omega = (-L_{x}/2,L_{x}) \times [-1,1] \times [0,L_{z}/2]
\,. 
\] 
Within this domain $\Omega$ there are then just four \stagp s. They 
are $SP_1$, $SP_3$, $SP_{N1}$, and $SP_{N2}$, shown in 
\reffig{fig:stagps_label2}. Note that $SP_{N2}$ is a translated 
version from the way it was viewed in \reffig{fig:stagps_label}. The 
phase portrait of fundamental dynamics for $EQ_2$ will be viewed in 
$\Omega$.

\subsection{A colorful flow portrait and {\hc}s}

With identification of all of the {\stagp}s within either the original 
periodic box or the cell $\Omega$, as well as the corresponding linear 
stability analysis, we are ready to make a complete phase space portrait 
for the upper branch, $EQ_2$.

\begin{figure}[!h]
\includegraphics[width=1.0\textwidth]{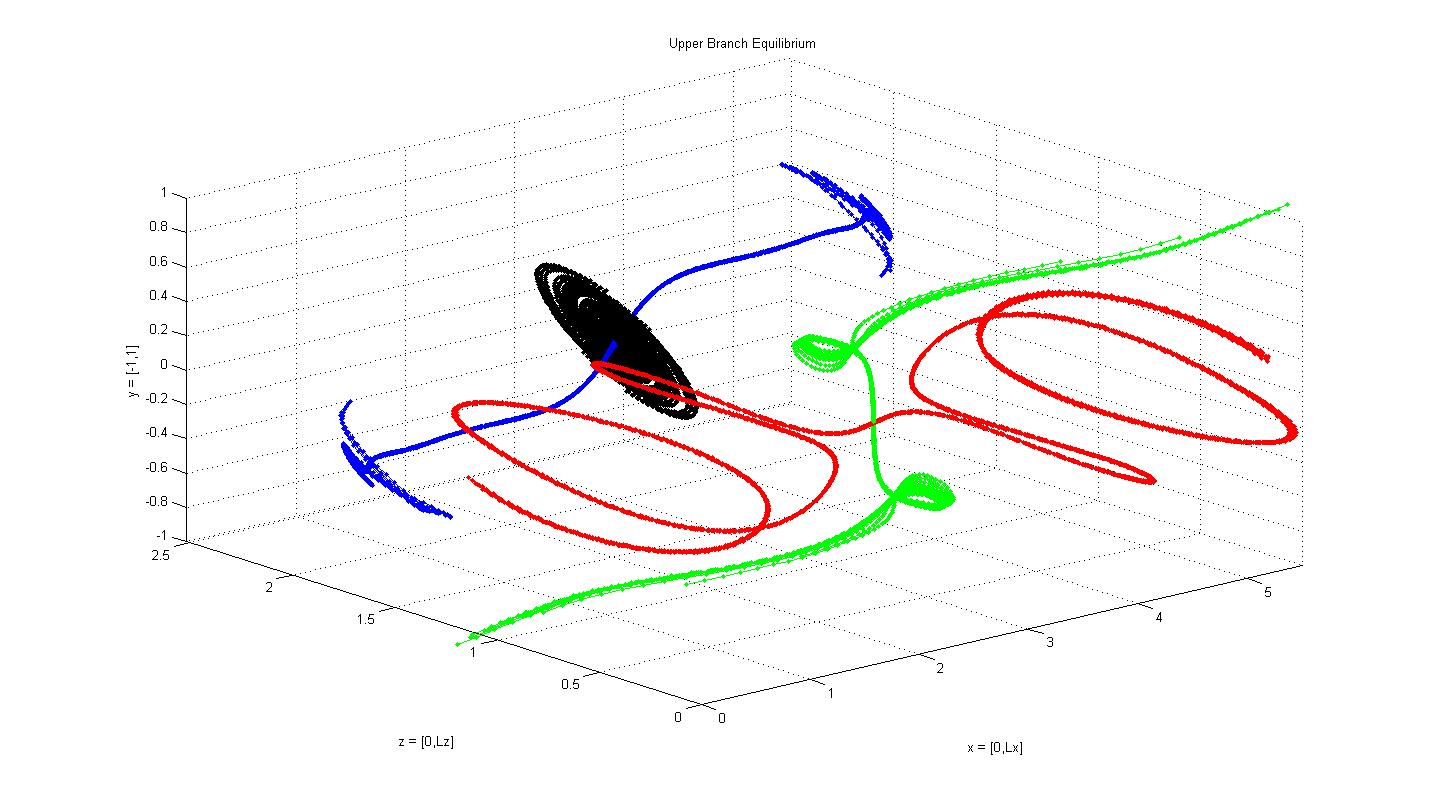}
  \caption{
   Segments of the stable (red/blue) and unstable (green/black) manifolds of the \stagp s
   $\bold{x}_{_{SP_{1}}} = (L_x/2,0,L_z/4)$ and
   $\bold{x}_{_{SP_{2}}} = (L_x/2,0,3L_z/4)$ for $EQ_2$. 
   }
  \label{fig:manifolds_both}
 \end{figure}

    \begin{figure}[!h]
\includegraphics[width=1.0\textwidth]{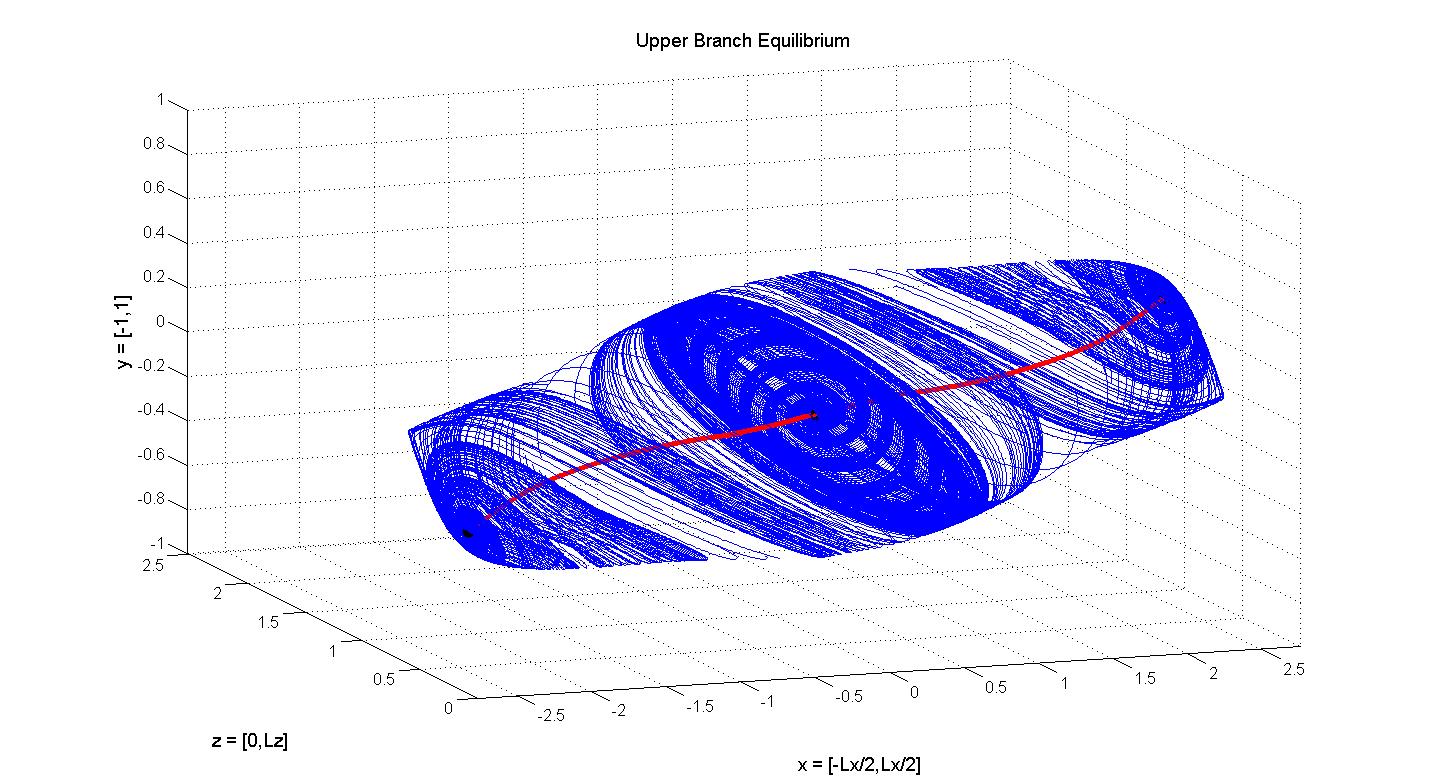}
  \caption{
{\Hec}s of the upper branch (red trajectories) from 
$SP_{N1}$ -> $SP_3$ and $SP_{N2}$ -> $SP_3$, shown in a cell with $x \in$ 
[-$L_x/2$, $L_x/2$] along with the unstable manifold of $SP_3$. 
   }
  \label{fig:hetero1}
 \end{figure}

  \begin{figure}[!h]
\includegraphics[width=1.1\textwidth]{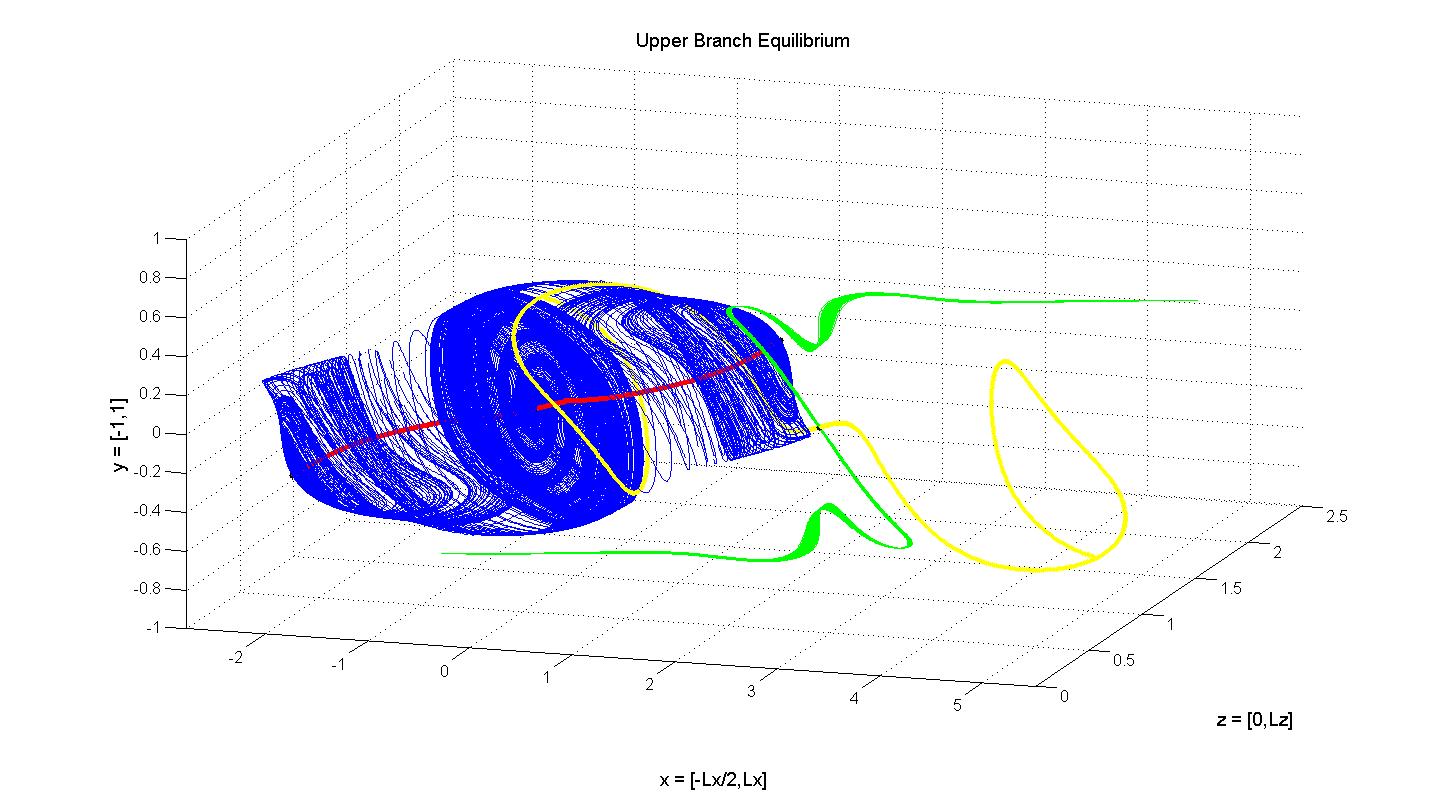}
  \caption{
Portrait of the fundamental dynamics along the manifolds of \stagp s 
$SP_1$, $SP_3$, $SP_{N1}$, $SP_{N2}$ within cell $\Omega$ for the upper 
branch. 
   }
  \label{fig:hetero2}
 \end{figure}

The dynamics between the \stagp s and their translations is quite 
interesting. In \reffig{fig:manifolds_both} we see a partial view of 
the stable and unstable manifolds of two of the {\stagp}s, $SP_1$ and 
$SP_2$, in the original periodic domain, found by integrating 
trajectories near the fixed points forwards and backwards in time along 
the stable or unstable eigenvectors. Local stability analysis shows that 
$SP_1$ has all real eigenvalues with a $1D$ stable manifold, and a $2D$ 
unstable manifold which is locally a plane 
\refeq{sp1_evec1}-\refeq{sp1_evec3}. The fact that one of the eigenvalues 
for the unstable manifold of $SP_1$ is much larger than the other is 
apparent in the figure by the fact that the trajectories in the unstable 
plane become quickly contracted in one of the dimensions, and the 
trajectories appear to leave along a nearly one-dimensional structure in 
the $y$-direction. $SP_2$ has a $2D$ unstable manifold with complex 
eigenvalues which spiral out in a plane and a $1D$ stable manifold. 
 
As alluded to in \reffig{fig:fine_usquare}, $SP_{N1}$ and $SP_{N2}$ 
sit near the center of the swirl of green coming from the unstable 
direction of $SP_1$. To better understand what is happening here, 
referring to \reffig{fig:hetero1}, we compute the  stable and 
unstable manifolds of $SP_{N1}$ and $SP_{N2}$, where we use the shifted 
translation of $SP_{N2}$, along with the stable and unstable manifolds of 
$SP_3$. The blue surface is formed by the overlap of trajectories 
starting along the unstable manifold of $SP_3$ and the stable manifolds 
of $SP_{N1}$ and $SP_{N2}$.  We see that the stable manifold of $SP_3$ 
(shown by the red curves) corresponds with the unstable manifolds of 
$SP_{N1}$ and $SP_{N2}$, thus we have \textit{{\hc}s} from $SP_{N1}$ --> 
$SP_3$ and $SP_{N2}$ --> $SP_3$! The thick appearance of the red curves 
is simply so that they can be seen within the blue surface. They are 
actually just a single trajectory. 
 
Next we bring trajectories originating near $SP_1$ into the picture to 
see how the manifolds of this {\stagp} connect with those in 
\reffig{fig:hetero1}, producing the full dynamical portrait within 
$\Omega$.  The result is shown in \reffig{fig:hetero2}. Compare to 
\reffig{fig:stagps_label2} to see the locations of the {\stagp}s. 
The relation of the stable manifold of $SP_1$ (yellow curve) and the 
trajectories that are driven away from $SP_1$ in the unstable direction 
(green) to those of the blue surface is quite interesting. These 
trajectories tightly hug the blue surface as they spiral around it, 
appearing to be shielded from entering the volume it encompasses. This 
could have significant implications for the consideration of fluid mixing 
within {\pCf}, perhaps showing that it is difficult to achieve a 
uniformly mixed space for this particular equilibrium; a blob of ink that 
starts outside of the blue surface may have a difficult time ever 
entering the region! 
 
One merely translates the image in \reffig{fig:hetero2} in the $x$ 
direction by an amount $L_{x}$ to give a complete picture in any periodic 
cell. The same picture will also occur symmetrically (translated by 
$L_{x}/2$ and $L_{z}/2$) in the left half of the box.

\subsection{Equilibrium $EQ_8$: Additional Symmetries}
\label{sect:EQ8}

Having analyzed the upper branch equilibrium $EQ_2$, we next look at 
$EQ_8$, another equilibrium velocity field of {\pCf} which exhibits 
turbulent behavior at a lower Reynolds number, 270.

\begin{figure}[!h]
\includegraphics[width=0.9\textwidth]{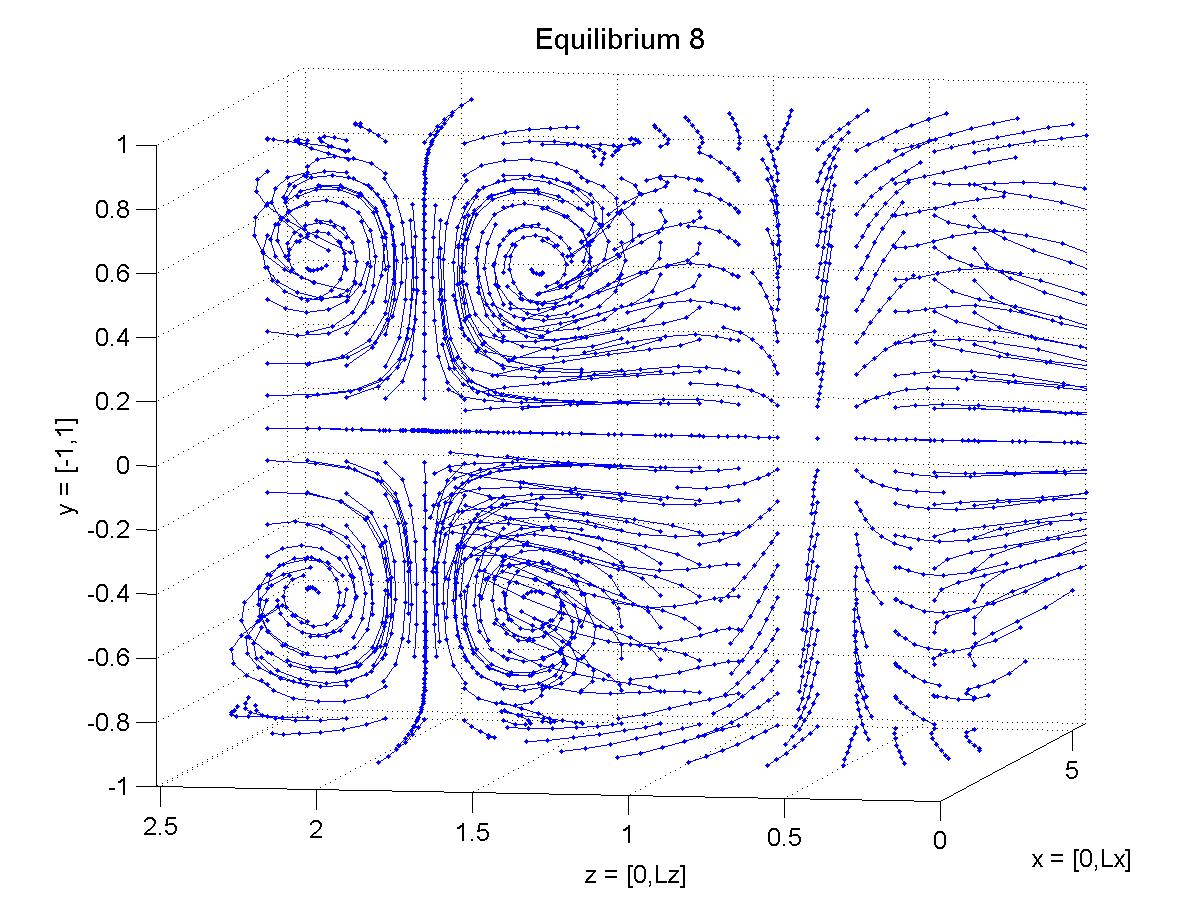}
  \caption{
    Grid of initial points in the $[y,z]$ plane, centered at $x = L_x/2$; 
    integrated to produce tracer particle trajectories for $EQ_8$. 
   }
  \label{fig:EQ8_grid1}
 \end{figure}

We start once again with a cleverly chosen grid of initial trajectories 
to get a feel for the significant structures in the flow. The grid is in 
a plane at $x = L_{x}/2$. The result, after a short integration time, is 
shown in \reffig{fig:EQ8_grid1}. This perspective view already shows 
us quite a bit of information. Once again we have symmetries abound, and 
we know from the discussion in \refsect{s:symm_stag} that there will be 
at least 8 {\stagp}s $SP_1$-$SP_8$.  Another interesting feature of this 
plot is the four vortical structures on the left half. One final 
noteworthy point from the figure is the appearance of a perfect line 
segment connecting two of the {\stagp}s, which happen to be $SP_1$ and 
$SP_2$. This strongly suggests a heteroclinic connection between these 
two \stagp s. To confirm, we compute the eigenvalues and eigenvectors of 
the \velgradmat. For $SP_1$, there is indeed a real, unstable eigenvector 
pointing along (0,0,1) and for $SP_2$ there is a real, stable eigenvector 
pointing along (0,0,1). This, together with the plot, numerically 
confirms the existence of the heteroclinic trajectory. The same result  
holds for the shifted pair at $x = 0$. The rest of the 
eigenvalues/eigenvectors are given below. We note that for $EQ_8$ there 
is a {\hc} which is a simple horizontal line connecting the pair of 
trivial \stagp s in the \textit{spanwise} direction, whereas for \tUB\ 
the connection was some arbitrary-looking curve in the 
\textit{streamwise} direction connected to a nontrivial \stagp. 
Factorization of the $SP_1$ and $SP_2$ stability eigenspaces for $EQ_8$ 
occurs because the spanwise $z$ direction is a $1D$ flow-invariant 
subspace at the \stagp s \cite{SiCvi10}. That ensures the simplicity of 
the \hec. 

$EQ_8$, $SP_1$: There are two real, positive eigenvalues
 and one real, negative eigenvalue.
\bea
\left(
    \eigExp[1],\eigExp[2],\eigExp[3]
\right) &=&
      (0.363557,0.227831,-0.591389)
\label{E8SP1} \\
\left(
    \jEigvec[1],\jEigvec[2],\jEigvec[3]
\right) &=&
\left(
    \begin{pmatrix}
             {0} \cr
             {0} \cr
             {1}
    \end{pmatrix} \,,
    \begin{pmatrix}
             {-0.733415} \cr
             {-0.679780} \cr
             {0}
    \end{pmatrix} \,,
    \begin{pmatrix}
             {0.991005} \cr
             {0.133824} \cr
             {0}
    \end{pmatrix}
\right) \,.
\nnu
\eea

$EQ_8$, $SP_2$: There are two real, positive eigenvalues
 and one real, negative eigenvalue.
\bea
\left(
    \eigExp[1],\eigExp[2],\eigExp[3]
\right) &=&
      (0.992857,0.255973,-1.248830)
\label{E8SP2} \\
\left(
    \jEigvec[1],\jEigvec[2],\jEigvec[3]
\right) &=&
\left(
    \begin{pmatrix}
             {~0.116961} \cr
             {-0.993136} \cr
             {0}
    \end{pmatrix} \,,
    \begin{pmatrix}
             {0.957795} \cr
             {0.287450} \cr
             {0}
    \end{pmatrix} \,,
    \begin{pmatrix}
             {0} \cr
             {0} \cr
             {1}
    \end{pmatrix}
\right) \,. \\
\nnu
\eea

Equilibrium $EQ_8$ (as well as $EQ_7$, not discussed here), possesses 
additional symmetries compared to $EQ_2$. $EQ_2$ is in the $S$-invariant 
subspace of velocity fields and $EQ_8$ is in $S_8$ (\refsect{s:PCF_symm}
and \refsect{s:symm_stag}). 

From \refeq{second_condition} and \refeq{s3lagrange} we know then that 
for $EQ_8$ we will have the additional {\stagp}s: 
 \bea
  \bold{x}_{_{SP_{5}}} &=& (L_x/4,0,0) \continue
  \bold{x}_{_{SP_{6}}} &=& (3L_x/4,0,0) \continue
  \bold{x}_{_{SP_{7}}} &=& (L_x/4,0,L_z/2)  \\
  \bold{x}_{_{SP_{8}}} &=& (3L_x/4,0,L_z/2) \nnu
 \,.
\eea
Interestingly these were actually discovered numerically \textit{before} 
the symmetry arguments were understood. A Newton search on regions of 
very low velocity for $EQ_8$ revealed that $(L_x/4,0,L_z/2)$ and 
$(3L_x/4,0,L_z/2)$ are \stagp s. From this, one may deduce that symmetry 
$s_5$ must hold, and it can then be checked that at any position the 
velocity field is indeed invariant under $s_4$ and $s_5$. 

Stability analysis of the additional set of {\stagp}s for $EQ_8$ gives the
following.

 $SP_5$: There is one real, positive eigenvalue
 and a complex pair with negative real part.
  \begin{align} &\eigExp[1] = 0.03109 \,,\quad \jEigvec[1] =
\begin{pmatrix}
             {0.85275} \cr
             {0.41774} \cr
             {-0.31355} \cr
   \end{pmatrix}
   \\
&\{ \eigExp[2],\eigExp[3]\}
  = \eigRe[2] \pm i \,\eigIm[2] =  -0.01555 \pm i\, 0.59385
   \label{EQSP5eigs}\\
&\jEigvec[2] =
\begin{pmatrix}
             {~0.24762} \cr
             {-0.31442} \cr
             {~0.69906} \cr
   \end{pmatrix}
    \,,\quad
\jEigvec[3] =
\begin{pmatrix}
             {-0.20793} \cr
             {~0.55489} \cr
             {~0} \cr
   \end{pmatrix}
\,.
\end{align}
 We have a $1D$ unstable manifold and a $2D$ inward-spiral
stable manifold. All four of the new points have the same
eigenvalues. $SP_5$ and $SP_8$ have the same eigenvectors, as do $SP_6$
and $SP_7$ whose eigenvectors differ from $SP_5$ only by the sign of
the third component for \jEigvec[1] and by the sign of the first and
second components for \jEigvec[2] and \jEigvec[3].

\begin{figure}[!h]
\centering
    \begin{subfigure}{0.9\textwidth}
    \includegraphics[width=1.0\textwidth]{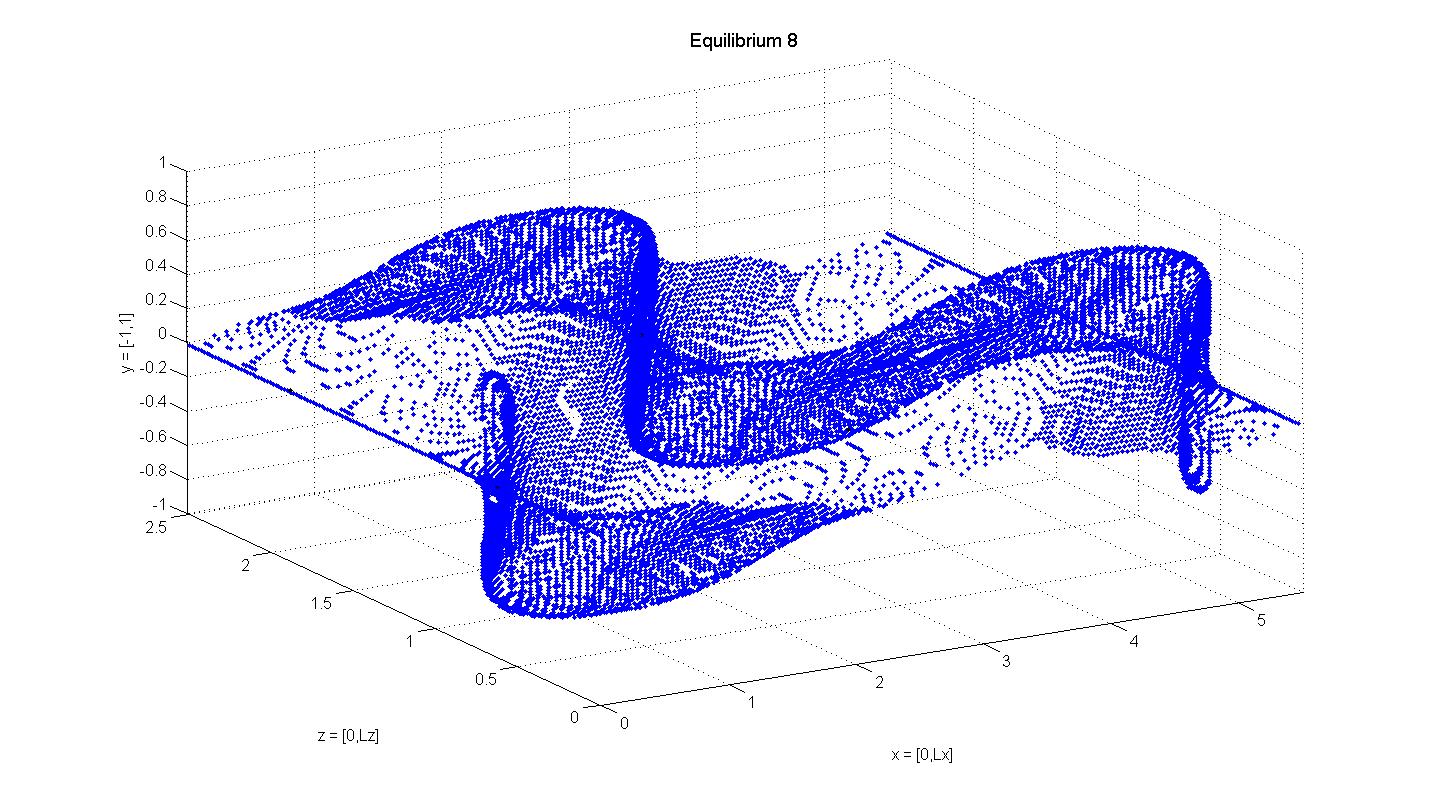}
      \caption{
        Perspective view.
       }
      \label{fig:usquare_EQ8_1}
    \end{subfigure}

    \begin{subfigure}{0.9\textwidth}
    \includegraphics[width=1.0\textwidth]{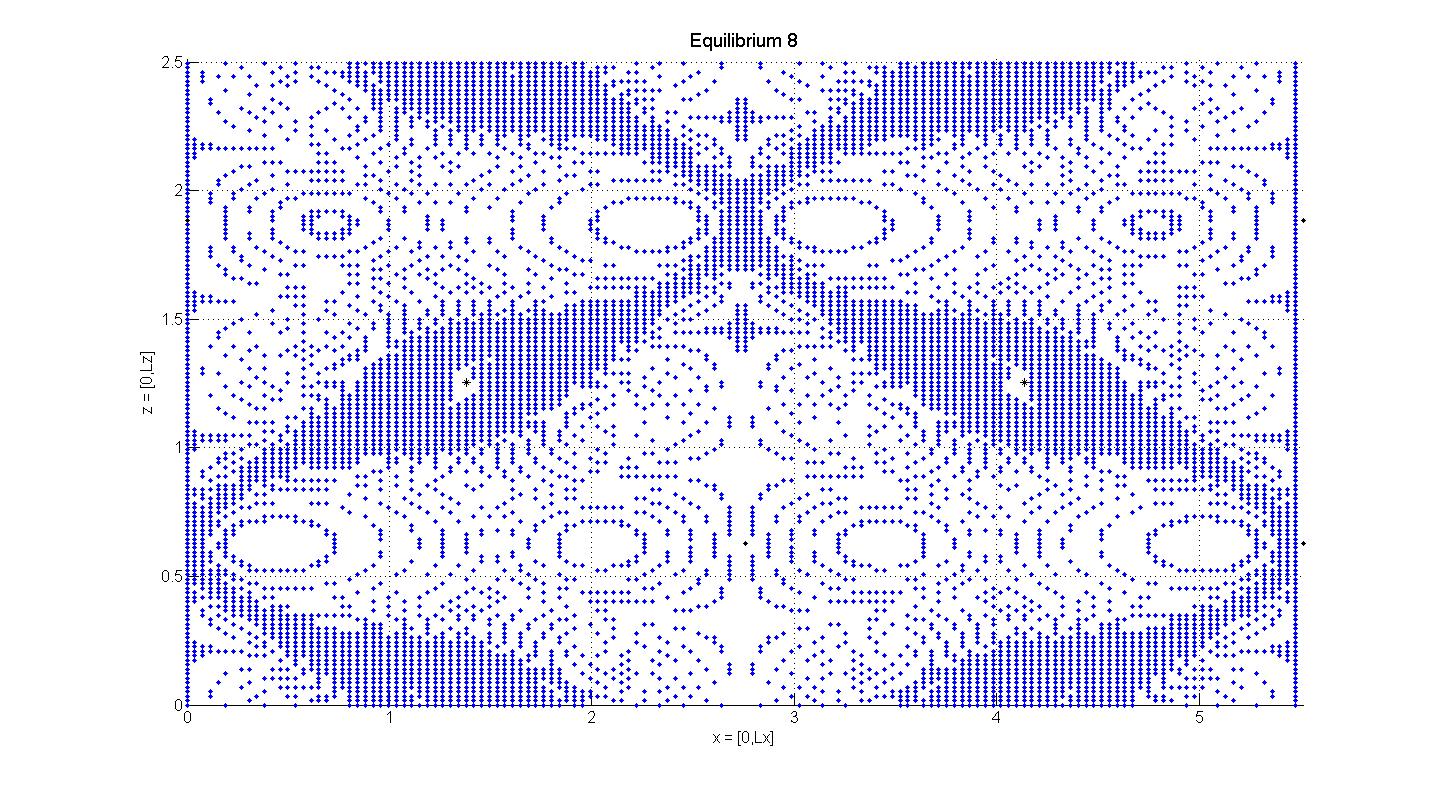}
      \caption{
       Projection onto the $xz$
       plane.
       }
      \label{fig:usquare_EQ8_2}
    \end{subfigure}  
    \caption{
A plot of points where the velocity field falls below a small cutoff for 
$EQ_8$, showing interesting structures in the flow. 
       }
    \label{fig:usquare_both}
 \end{figure}

As a final interesting consequence of numerically searching for \stagp s 
for $EQ_8$, the figures produced by plotting gridpoints where velocity is 
small, using a cutoff value of $|\mathbf{u}|^{2}$ which is too large to 
actually be useful for finding \stagp s, we instead find a plot showing 
more intricate patterns in the flow. \reffig{fig:usquare_EQ8_1} 
shows a $3D$ perspective view of these points, and 
\reffig{fig:usquare_EQ8_2} shows the projection of 
\reffig{fig:usquare_EQ8_1} onto the $xz$ plane. This 
volume-preserving flow (area preserving in Poincar\'e sections) may have 
invariant tori which, being quasiperiodic, would not be detected by the 
{\stagp} searching routines. Though the structures in the 
projection plot in \reffig{fig:usquare_EQ8_2} are not actual tracer 
trajectories, they are suggestive that a search for such invariant tori 
in future work may be a fruitful endeavor.  

\section{Conclusion}
\label{s:conclusion}

We have taken a step towards a deeper understanding of the  turbulent 
fluid flow in a $3D$ system from the Lagrangian perspective by studying 
tracer trajectory dynamics in plane Couette geometry. Potential 
applications that could follow from having a grasp of the Lagrangian 
dynamics and being able to accurately compute tracer particle 
trajectories are wide-ranging: velocity profile statistics or correlation 
functions taken over an ensemble of particle trajectories within 
different regions, calculations of mixing time and diffusion properties 
for the flow, Lyapunov exponents and material stretching, striation 
thickness, among others, are some of the various possible measures of 
chaotic advection that could be investigated. By extending the dynamical 
systems methods that are often confined to simpler $2D$ systems to the 
$3D$ world of {\pCf}, we encounter complex coherent structures that 
partition the physical space of the fluid into regions which exhibit 
distinct types of motion and allow us to visualize the fundamental 
motions driven by trajectories which lie close to invariant manifolds. 
Relying on the symmetries of the geometry to shine light upon the 
situation and guide us, we are able to construct phase portraits for 
plane Couette equilibria starting with the identification and stability 
determination of stagnation or fixed points of the system. Future work 
could easily extend these analyses to additional invariant solutions for 
{\pCf}, or apply the same methods in other fluid systems which likely 
posses symmetries. 

\bibliographystyle{unsrt}
\bibliography{../../../bibtex/pipes} 
\end{document}

%% file: defsElton.tex
\newsavebox{\bartName}

{\hspace*{\fill}\nolinebreak[1]\usebox{\bartName}\vspace*{1ex}\end{minipage}}
\newcommand{\beq}{\begin{equation}}
\newcommand{\continue}{\nonumber \\ }
\newcommand{\nnu}{\nonumber}
\newcommand{\eeq}{\end{equation}}
\newcommand{\ee}[1] {\label{#1} \end{equation}}
\newcommand{\bea}{\begin{eqnarray}}

\newcommand{\eea}{\end{eqnarray}}
\newcommand{\barr}{\begin{array}}
\newcommand{\earr}{\end{array}}

\newcommand{\refeq}  [1] {(\ref{#1})}

\newcommand{\reffig} [1] {figure~\ref{#1}}

\newcommand{\refsect}[1] {sect.~\ref{#1}}

\newcommand{\statesp}{state space}

\newcommand{\BER}[1]{{\mbox{\footnotesize BER}}} 

\newcommand {\id}{{\ \hbox{{\rm 1}\kern-.62em\hbox{\rm 1}}}} 

\newcommand{\stagp}{stagnation point}

\newcommand{\velgradmat}{velocity gradients matrix}

\newcommand{\Mvar}{\ensuremath{A}}  
\newcommand{\jEigvec}[1][]{\ensuremath{{\bf e}^{(#1)}}} 

\newcommand{\eigExp}[1][]{
     \ifthenelse{\equal{#1}{}}{\ensuremath{\lambda}}{\ensuremath{\lambda^{(#1)}}}}
\newcommand{\eigRe}[1][]{
     \ifthenelse{\equal{#1}{}}{\ensuremath{\mu}}{\ensuremath{\mu^{(#1)}}}}
\newcommand{\eigIm}[1][]{
     \ifthenelse{\equal{#1}{}}{\ensuremath{\omega}}{\ensuremath{\omega^{(#1)}}}}

\newcommand{\eqv}{equi\-lib\-rium}

\newcommand{\eqva}{equi\-lib\-ria}

\newcommand{\reqva}{rela\-ti\-ve equi\-lib\-ria}
\newcommand{\Hec}{Hetero\-clinic connect\-ion}
\newcommand{\hec}{hetero\-clinic connect\-ion}

\newcommand{\shift}{\ensuremath{d}}

\newcommand{\bbU}{\mathbb{U}}
\newcommand{\bbUsymm}{\bbU_{c}}

\newcommand{\hc}{heteroclinic connection}

\newcommand{\NS}{Navier-Stokes}
\newcommand{\NSe}{Navier-Stokes equation}
\newcommand{\pCf}{plane Couette flow}
\newcommand{\PCf}{Plane Couette flow}
\newdimen\onebox
\newdimen\boxsize
\newcount\boxnum
\gdef\mult#1#2#3{
    \ifx#1\relax\else%
      \ifx#2\relax\else%
        #1=#2%
        \ifx#3\relax\else%
          \multiply#1#3%
        \fi%
      \fi%
    \fi}

\newlength{\verti}

\thicklines
\newlength{\Fsize}   
\newlength{\Fdotsize}
\newcommand{\Reynolds}{\ensuremath{\textit{Re}}}  

\newcommand{\grad}{{\bf \nabla}}

\newcommand{\bu}{\ensuremath{{\bf u}}}
\newcommand{\bx}{\ensuremath{{\bf x}}}

\newcommand{\butot}{\ensuremath{{\bf u_{tot}}}}
\newcommand{\bnabla}{\ensuremath{{\bf \nabla}}}
\newcommand{\lapl}{\ensuremath{{\nabla^{2}}}}

\newcommand{\tUB}{\ensuremath{{\text{EQ2}}}}

